\documentclass[lettersize,journal]{IEEEtran}
\usepackage{amsmath,amsfonts}
\usepackage{mathtools} 
\usepackage{algorithm}
\usepackage{algpseudocode}
\usepackage{array}
\usepackage{subcaption}
\usepackage{textcomp}
\usepackage{stfloats}
\usepackage{url}
\usepackage{verbatim}
\usepackage{graphicx}
\usepackage{float}
\usepackage{cite}
\usepackage{orcidlink}
\usepackage{multirow}

\begin{document}

\title{Deep Fuzzy Framework for Emotion Recognition using EEG Signals and Emotion Representation in Type-2 Fuzzy VAD Space}


\author{Mohammad Asif \orcidlink{0000-0002-9517-6716},~\IEEEmembership{Graduate Student Member,~IEEE}, 
        Noman Ali \orcidlink{0000-0002-7315-6984}, Sudhakar Mishra \orcidlink{0000-0002-3748-7153}, Anushka Dandawate \orcidlink{0009-0006-2682-0092}, Uma Shanker Tiwary \orcidlink{0000-0001-7206-9013},~\IEEEmembership{Senior Member,~IEEE.}
\thanks{Manuscript received April 19, 2021; revised August 16, 2021. \textit{(Corresponding author: Mohammad Asif)} \\
Mohammad Asif, Uma Shanker Tiwary, and Anushka Dandawate are with the Indian Institute of Information Technology Allahabad, Prayagraj, 211012, India (email: pse2017001@iiita.ac.in; ust@iiita.ac.in; iit2020077@iiita.ac.in).

Noman Ali is with the Indian Institute of Technology Madras, Chennai 600036, India, and also with Panjab University, Chandigarh 160014 (email: 21f1004271@ds.study.iitm.ac.in, alinoman.itz@gmail.com ).

Sudhakar Mishra is with the Indian Institute of Technology Kanpur, Kanpur 208016, India (email: sudhakarm@iitk.ac.in).
}}

\markboth{Journal of \LaTeX\ Class Files,~Vol.~14, No.~8, August~2021}%
{Shell \MakeLowercase{\textit{et al.}}: A Sample Article Using IEEEtran.cls for IEEE Journals}


\maketitle

\begin{abstract}
Recently, the representation of emotions in the Valence, Arousal and Dominance (VAD) space has drawn enough attention. However, the complex nature of emotions and the subjective biases in self-reported values of VAD make the emotion model too specific to a particular experiment. This study aims to develop a generic model representing emotions using a fuzzy VAD space and improve emotion recognition by utilizing this representation. We partitioned the crisp VAD space into a fuzzy VAD space using low, medium and high type-2 fuzzy dimensions to represent emotions. A framework that integrates fuzzy VAD space with EEG data has been developed to recognize emotions. The EEG features were extracted using spatial and temporal feature vectors from time-frequency spectrograms, while the subject-reported values of VAD were also considered. The study was conducted on the DENS dataset, which includes a wide range of twenty-four emotions, along with EEG data and subjective ratings. The study was validated using various deep fuzzy framework models based on type-2 fuzzy representation, cuboid probabilistic lattice representation and unsupervised fuzzy emotion clusters. These models resulted in emotion recognition accuracy of 96.09\%, 95.75\% and 95.31\%, respectively, for the classes of 24 emotions. The study also included an ablation study, one with crisp VAD space and the other without VAD space. The result with crisp VAD space performed better, while the deep fuzzy framework outperformed both models. The model was extended to predict cross-subject cases of emotions, and the results with 78.37\% accuracy are promising, proving the generality of our model. The generic nature of the developed model, along with its successful cross-subject predictions, gives direction for real-world applications in the areas such as affective computing, human-computer interaction, and mental health monitoring.
\end{abstract}

\begin{IEEEkeywords}
EEG Emotion Recognition, Affective Computing, VAD, Emotion Dimensions, DENS Dataset, Emotion Analysis, Emotion Representation, CNN, LSTM, Multimodal Fusion, Deep Fuzzy Framework, Emotion Modelling.
\end{IEEEkeywords}

\section{Introduction}
\begin{figure}[!t]
\centering
\includegraphics[width=0.8\linewidth]{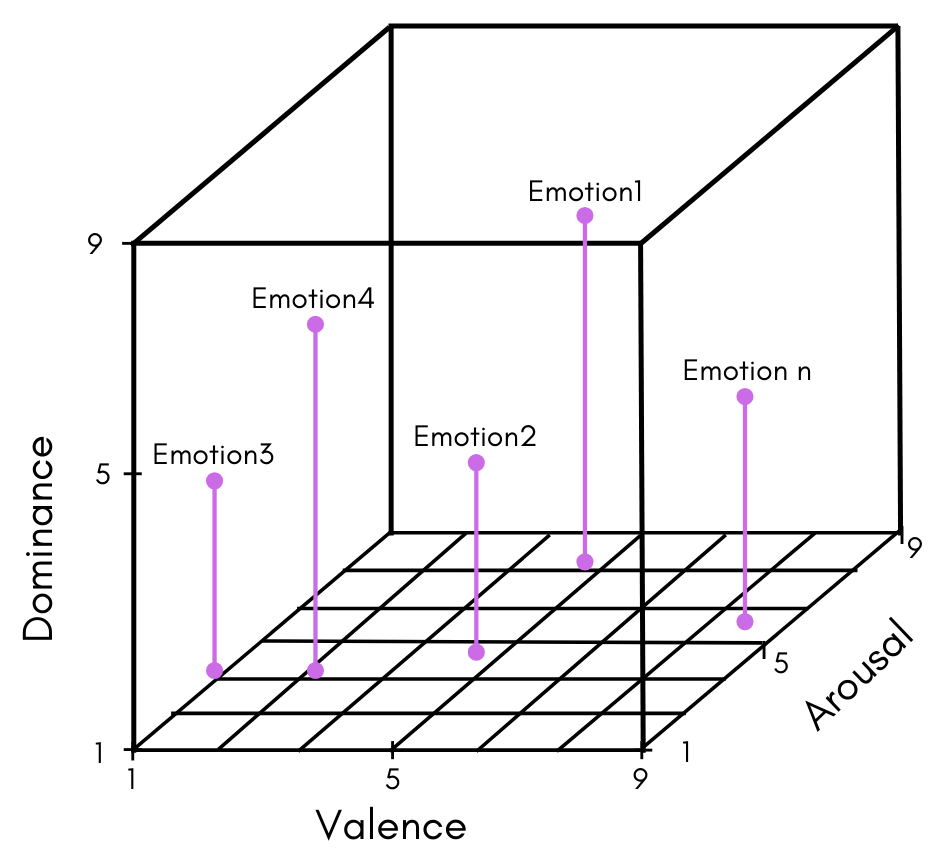}
\caption{Emotion Dimensions: A schematic diagram depicting emotion in VAD space.}
\label{fig:VAD}
\end{figure}

\IEEEPARstart{I}{dentifying} and understanding emotions has been a major challenge not only in the field of artificial intelligence but also in general science and psychology. We all experience a range of emotions in our daily lives. It is a complex phenomenon. According to the multidimensional theory of emotions, these psychological phenomena are complex constructs that consist of different dimensions\cite{kesebir2019emotional}, which together shape the human emotional experience. These multidimensional spaces are Valence, Arousal and Dominance (VAD). In this perspective, the complexity of emotions arises from the interplay of multiple dimensions. Basic emotions can be considered as primary building blocks within a multidimensional space\cite{saariluoma2014emotional}. More subtle and complex emotions emerge through the blending, intensifying, or suppressing of these basic emotional dimensions\cite{ekman1992there, griffiths2002basic, ortony2022all}.  A schematic depiction of the emotion dimensions is provided in Fig.\ref{fig:VAD}. We can depict the emotion-\(i\) representation in VAD space by
\begin{equation}\label{eq:crispFn}
    E_i \sim f(x_V,x_A,x_D)
\end{equation}
where \(E\) $\in$ \{\text{Set of Emotions}\}, \(f\) represents a function, and \( x_V,x_A,x_D \) are the crisp values of valence, arousal and dominance provided by the participants. 

Emotions are not simple entities but complicated combinations of subjective feelings, physiological responses, cognitive evaluations, and behavioural tendencies\cite{barrett2017emotions}. Moreover, the temporal dynamics of emotions should not be examined as isolated events but as evolving trajectories within this multidimensional framework. The subjective experience of an emotion is closely linked to the cognitive appraisal of a situation, the physiological arousal it induces, the subsequent behavioural expressions, and the element of perceived control or influence within the emotional experience\cite{barrett2017emotions}. This study also aims to explore the importance of a multidimensional theory and why valence, arousal, and dominance are important in understanding a wide spectrum of emotions. A special emphasis will be placed on Dominance and empirically showing why it is so important while considering a complex phenomenon like emotions and how it is related to the user's subjective experience for controlling the emotions.

Human emotions frequently involve a degree of ambiguity, where individuals may experience mixed feelings or transitions between emotional states\cite{braniecka2014mixed}. Also, emotional expression and perception vary across cultures and individuals. Individuals vary in how they express and experience emotions. Some may express joy more exuberantly, while others may do so more subtly. Emotional intensity often varies gradually rather than abruptly\cite{van2023modelling}. Also, due to a limited emotional vocabulary, it is challenging to convey the nuances of their feelings accurately. The lack of precise words for certain emotional states contributes to ambiguity in self-reports. The meanings of emotional terms can vary among individuals. What one person labels as "anxiety," another might perceive as "nervousness" or "apprehension". This semantic variability introduces ambiguity in the interpretation of self-reports. This ambiguity also occurs from the individuals' appraisal and interpretation of emotional stimuli differently as they may use different terms to describe similar emotional responses based on their cognitive interpretations. Due to all these and because human emotions are inherently complex and often exist along a spectrum\cite{van2022social}, crisp boundaries, being discrete and predefined, struggle to capture the full spectrum of emotional experiences. Emotions can manifest with varying intensities, shades, and combinations. For example, emotions like "mixed feelings" or "ambivalence" are challenging to represent within the confines of crisp categories. Individuals rarely experience emotions as isolated events but rather undergo fluid transitions from one emotional state to another\cite{li2020roles}. Fuzzy logic excels in handling such cases, and it can handle complex emotion combinations by allowing overlapping memberships. This flexibility enables the representation of mixed emotions or intricate blends, acknowledging that individuals often experience a combination of feelings simultaneously.
\linebreak

\noindent {\bf{EEG Based Emotion Recognition}}

Emotion recognition through EEG is a growing field in affective computing that involves studying the complex neural patterns associated with human emotions by examining brain activity during emotional experiences. EEG, a non-invasive method, offers a real-time window with high temporal precision, providing a valuable approach in diverse applications. This can range from improving human-computer interaction to contributing to mental health research and finding neural correlates of the brain for emotions and processing. 

EEG data analysis includes frequency bands, spectral power analysis, ERPs, connectivity measures, and time-frequency analysis. Emotion recognition is modelled using several methods, such as machine learning, ensemble learning, fuzzy logic, deep learning, and so on. Researchers use various models such as CNN, CNN-LSTM, CNN-GRU, attention-based hierarchical, and multi-head self-attention for emotion recognition \cite{zheng2015investigating, thammasan2016application, salama2018eeg, sheykhivand2020recognizing, 10101783, li2021hierarchical, hu2022eeg}.

However, there are challenges in this field, specifically defining and categorising emotions has always been a challenging task, including the complexity of emotions and the need to interpret intricate neural signals accurately\cite{10101783}. EEG-based emotion recognition provides an objective measure of emotional states, minimizing reliance on subjective self-reports. This objectivity is essential for avoiding biases and gaining a more accurate understanding of emotional responses. Additionally, there are other challenges that need to be overcome, including individual variability and integration with multimodal data like ECG, EMG, EOG and self-reported states \cite{hu2022eeg}. We will address and incorporate this issue in this study.
\linebreak

\noindent {\bf{Research Gap-Mapping The Relationship Between VAD Space and Emotion:}}

In emotion studies, prevailing research often focuses on discrete emotion recognition\cite{gannouni2021emotion} or continuous emotional dimensions recognition\cite{salankar2021emotion} and usually ignores the complex relationship between emotions and emotional dimensions. This study attempts to address this gap by incorporating the deep fuzzy framework grounded in a multidimensional VAD space of emotions. This study attempts to map the relationship between emotions and the dimensional space of emotions. The primary objective is to highlight the significance of multidimensions and emphasise why it is imperative not to overlook them when delving into the identification of complex emotions beyond discrete categories. Building upon the multidimensional theory of emotions, our approach acknowledges that emotions are multifaceted constructs encompassing various dimensions\cite{barrett2017emotions}.

Another significant challenge in this field is the availability of datasets based on EEG. It is constrained in terms of quantity. Generating the required amount of datasets based on EEG or any other biomedical imaging is not an easy task. Therefore, managing and optimising a limited dataset provides a significant challenge. A deep learning model often requires a substantial amount of data to be effectively trained, which is not possible with a limited dataset. However, utilising all the information inherited in the datasets can help to cope with this problem. EEG is one of the features of the datasets that can be used, but generally, the ratings provided by the participants are omitted. In this work, we aim to overcome this issue by considering information about emotional experience in terms of three independent dimensions - Valence, Arousal, and Dominance (VAD) with a deep fuzzy framework consisting of fuzzy VAD space. Fuzzy logic can be seamlessly integrated with deep learning models\cite{sorkhi2022hybrid}, combining the strengths of both paradigms. This integration allows for the development of deep fuzzy frameworks that leverage the power of neural networks while benefiting from the flexibility and interpretability of fuzzy logic\cite{talpur2022comprehensive}.
\linebreak

\noindent {\bf{Our Contributions:}}

Before going further, here is a brief mention of our contribution to this study in a nutshell:
\begin{itemize}
    \item An insight into a novel representation of emotions using type-2 fuzzy VAD space.
    \item \textbf{Deep Fuzzy Framework:} Deep Fuzzy Framework: Proposing a multimodal fusion framework for recognition of a wide spectrum of emotions using fuzzy VAD space and EEG data.
    \item Significantly improving cross-subject emotion recognition with the integration of Deep Fuzzy Framework.
    \item Representation of temporally localised twenty-four emotions (in DENS Dataset), developed and validated on Indian population in our lab\cite{10101783, mishra2021dataset}.
\end{itemize}

The paper is structured into several sections, starting with an introduction to the research topic. The following section covers the VAD space and provides an overview of the emotions that are included in it using the DENS dataset. The third section talks about the fuzzy representation of emotion. The fourth section is about the methodology, which discusses the complete methodology from the preprocessing of the EEG signal to the integration of fuzzy framework in the deep architecture. The next section presents the results of all the experiments done. Finally, the study concludes with a discussion in the last section.

\begin{figure*}[h!]
\centering
\includegraphics[width=\linewidth]{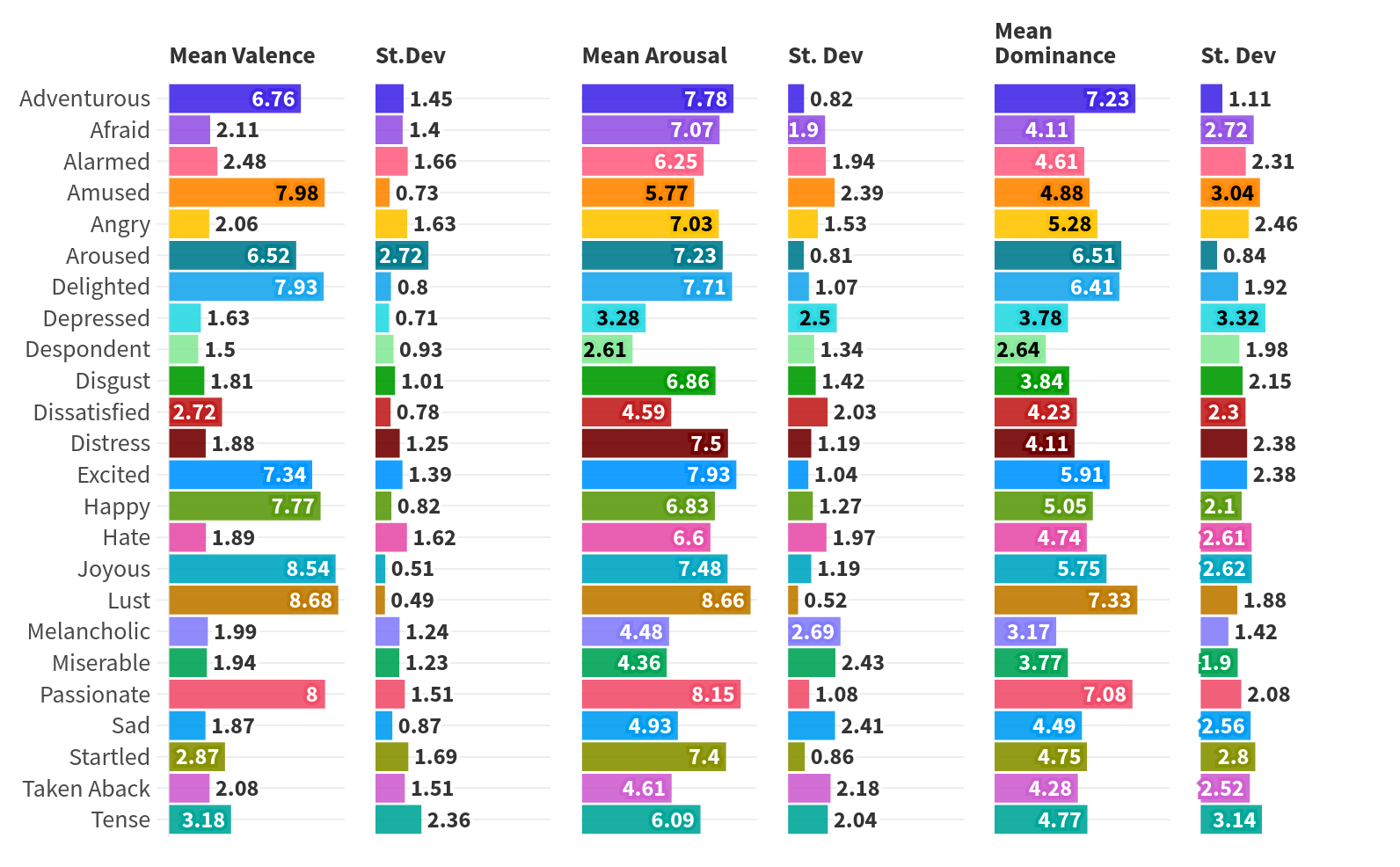}
\caption{Emotion Distribution: Mean and Standard Deviation of all the emotions.}
\label{fig:emotion_VAD_chart}
\end{figure*}

\begin{figure*}[h!]
\centering
 \begin{subfigure}{0.45\textwidth}
    \includegraphics[width=\linewidth]{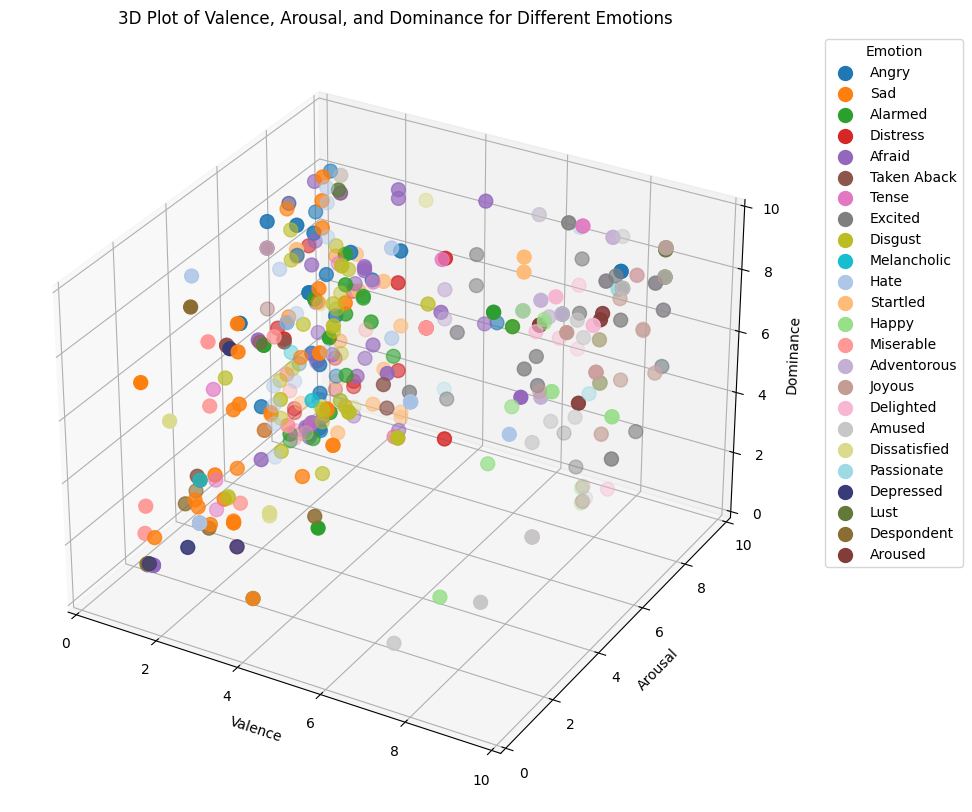}
    \caption{Distribution of All Emotions in VAD Space}
    \label{fig:sub1_Distribution}
  \end{subfigure}
  \hfill
  \begin{subfigure}{0.45\textwidth}
    \includegraphics[width=\linewidth]{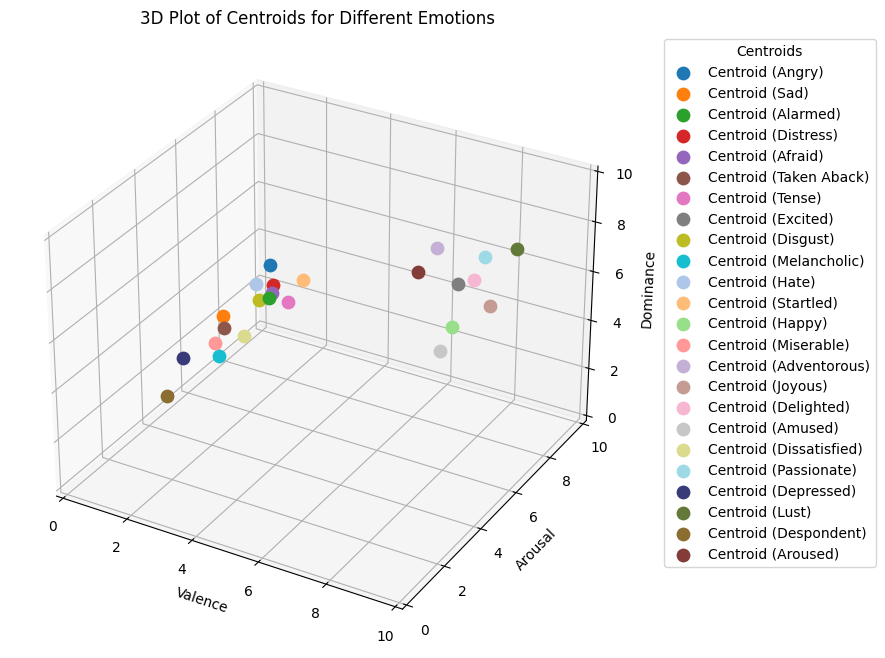}
    \caption{Location of Centroids in Each Emotion}
    \label{fig:sub2_Distribution}
  \end{subfigure}
\caption{Emotion Distribution of DENS}
\label{fig:Distribution}
\end{figure*}

\begin{figure*}[h!]
\centering
\includegraphics[width=\linewidth]{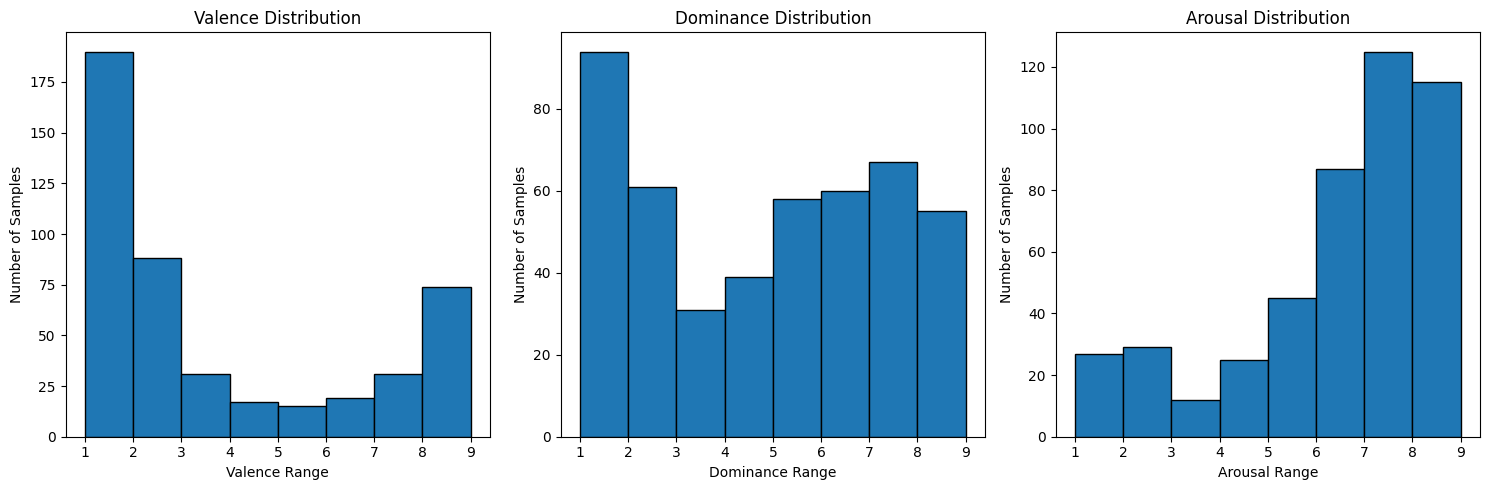}
\caption{VAD Distribution: A histogram of all emotional events in VAD space.}
\label{fig:VAD_distribution}
\end{figure*}

\begin{figure}[h!]
\centering
\includegraphics[width=\linewidth]{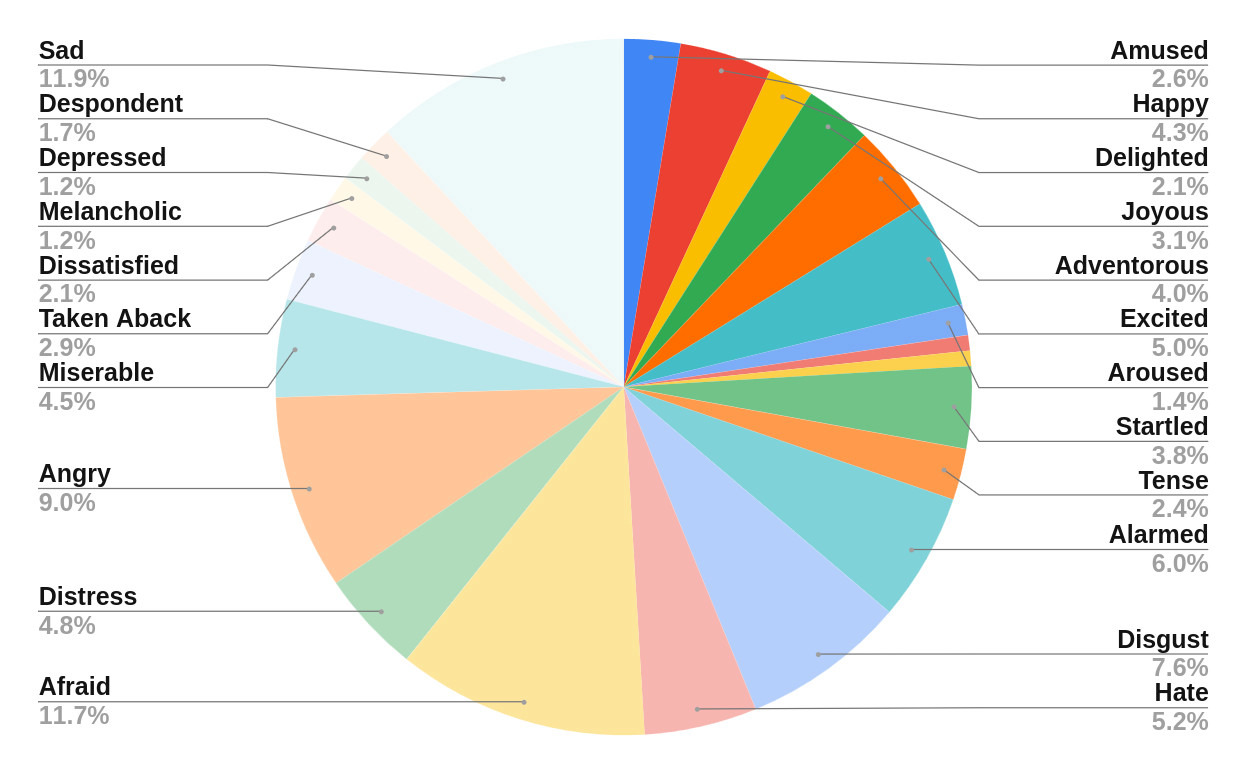}
\caption{Emotion Percent Distribution: Emotional Events Pie Chart.}
\label{fig:emotion_wheel}
\end{figure}

\begin{figure*}[h!]
\centering
 \begin{subfigure}{0.49\textwidth}
    \includegraphics[width=\linewidth]{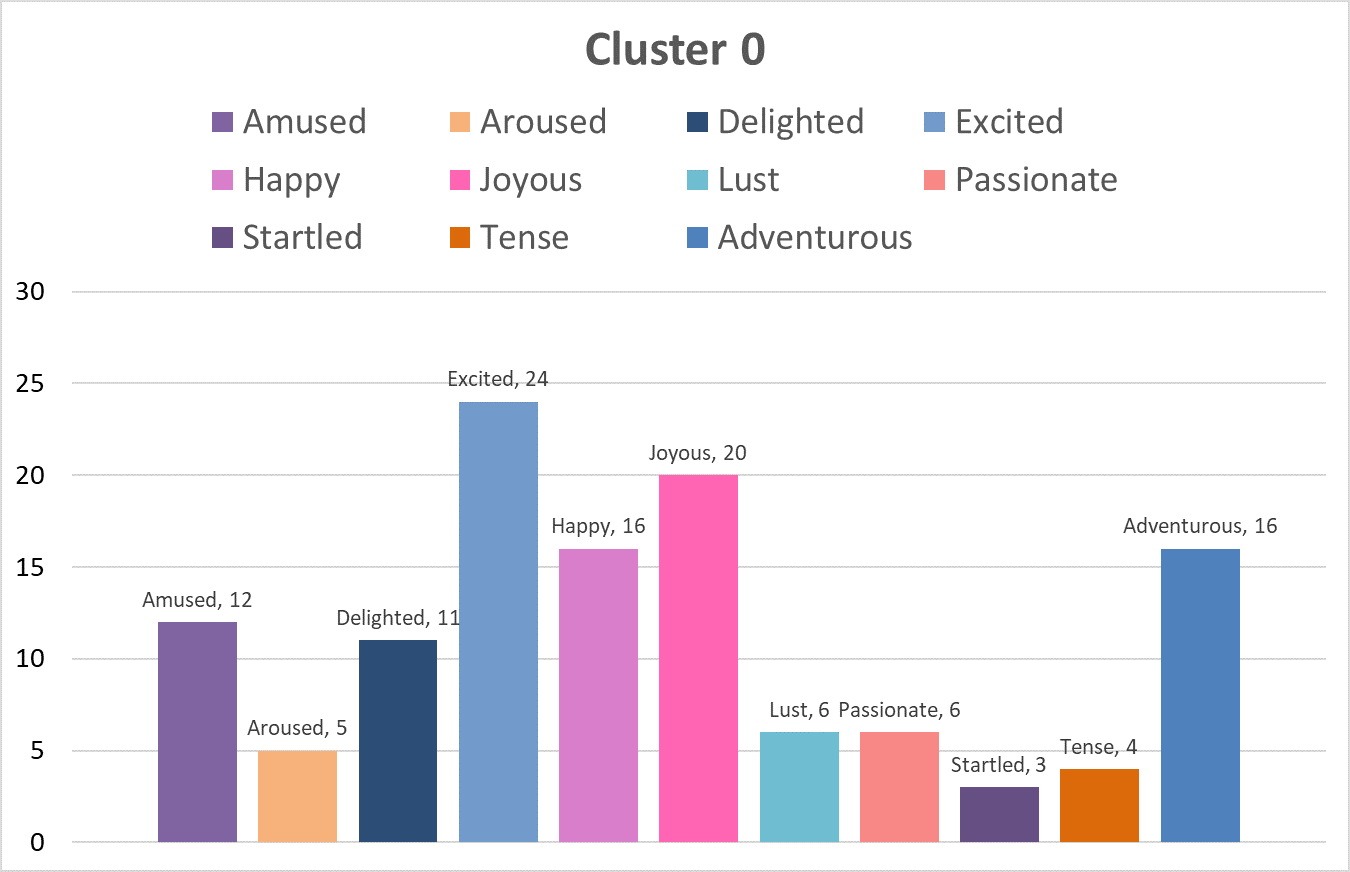}
  \end{subfigure}
  \hfill
  \begin{subfigure}{0.49\textwidth}
    \includegraphics[width=\linewidth]{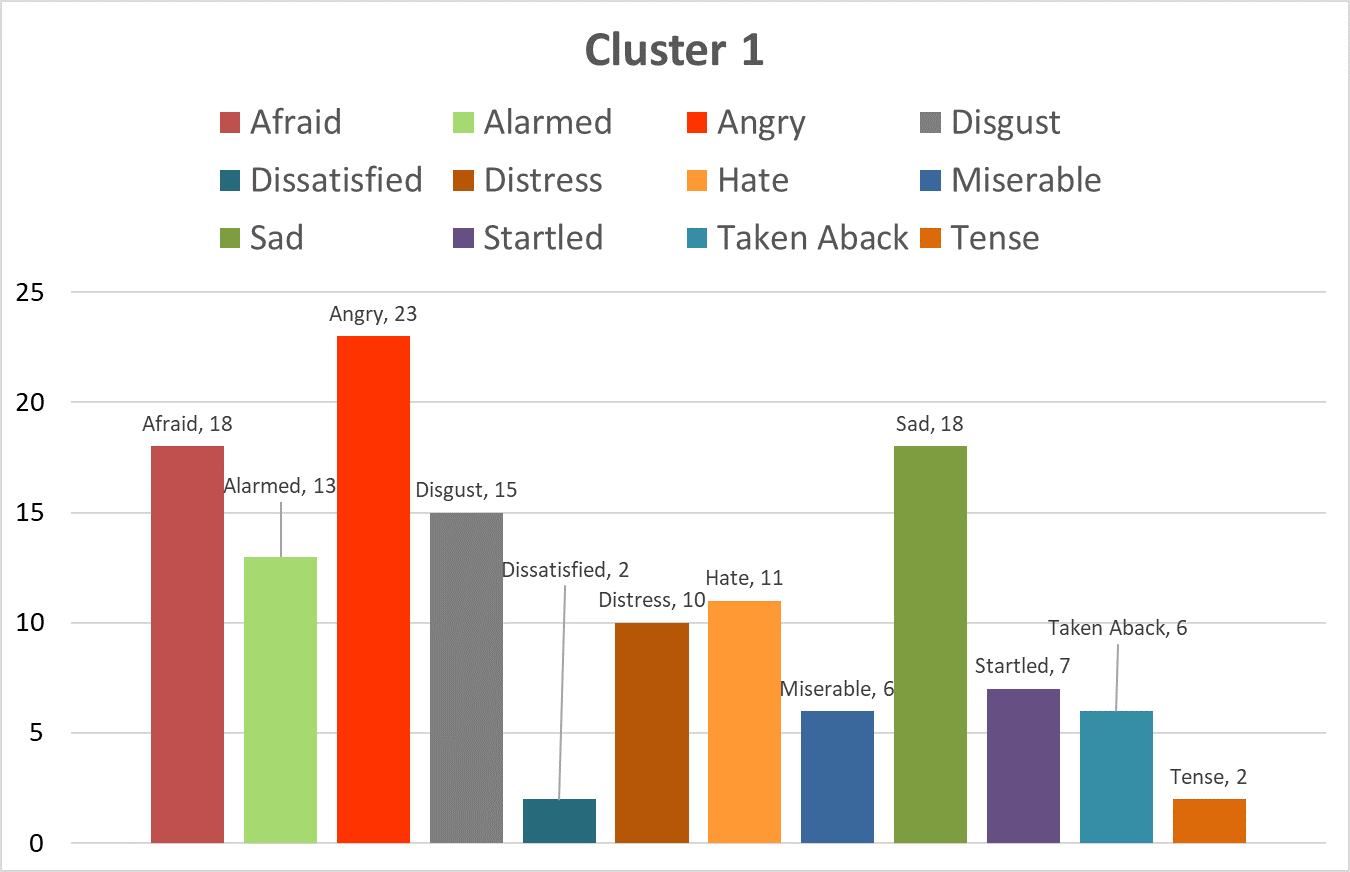}
  \end{subfigure}
 \begin{subfigure}{0.49\textwidth}
    \includegraphics[width=\linewidth]{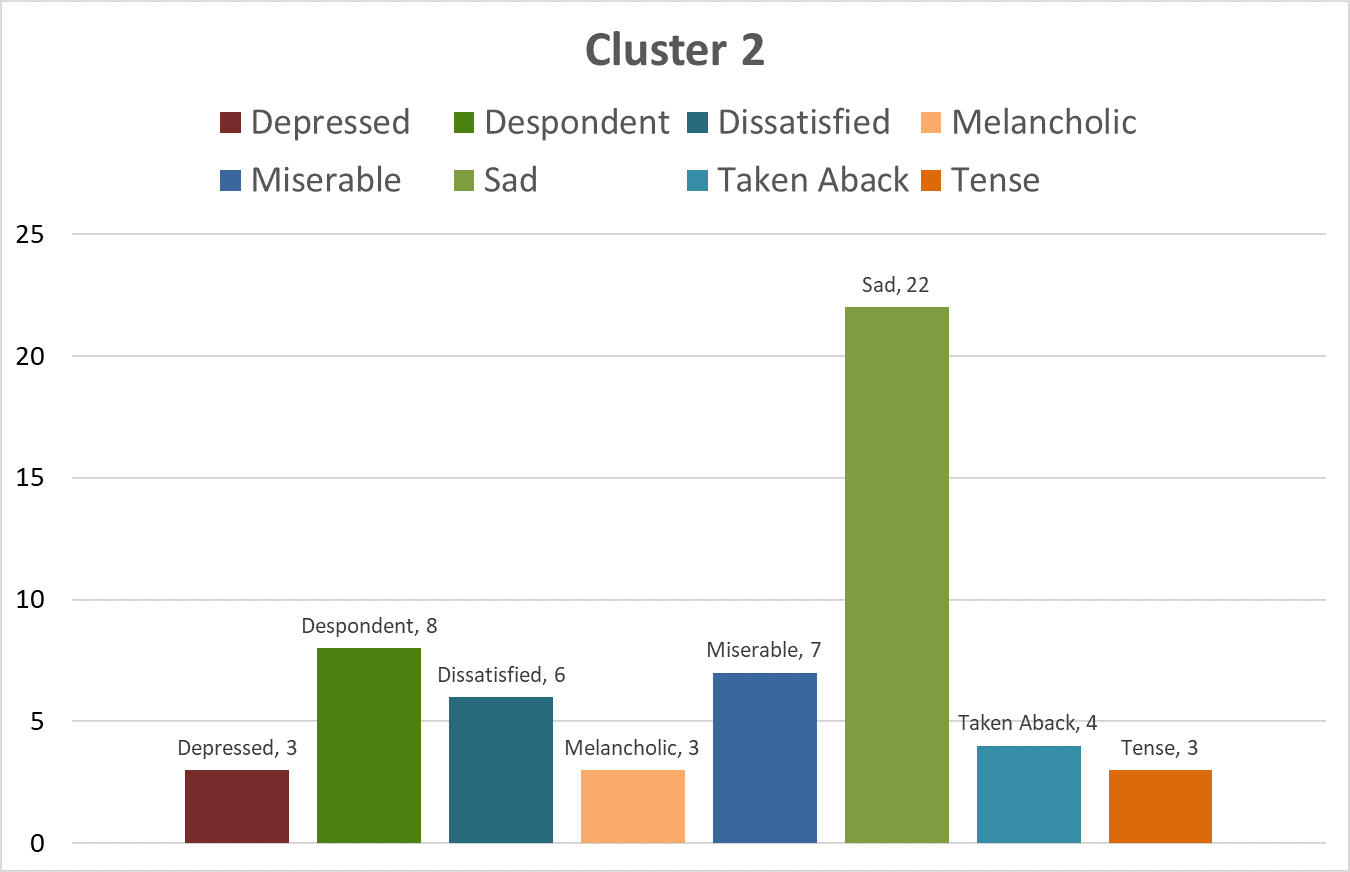}
  \end{subfigure}
  \hfill
  \begin{subfigure}{0.49\textwidth}
    \includegraphics[width=\linewidth]{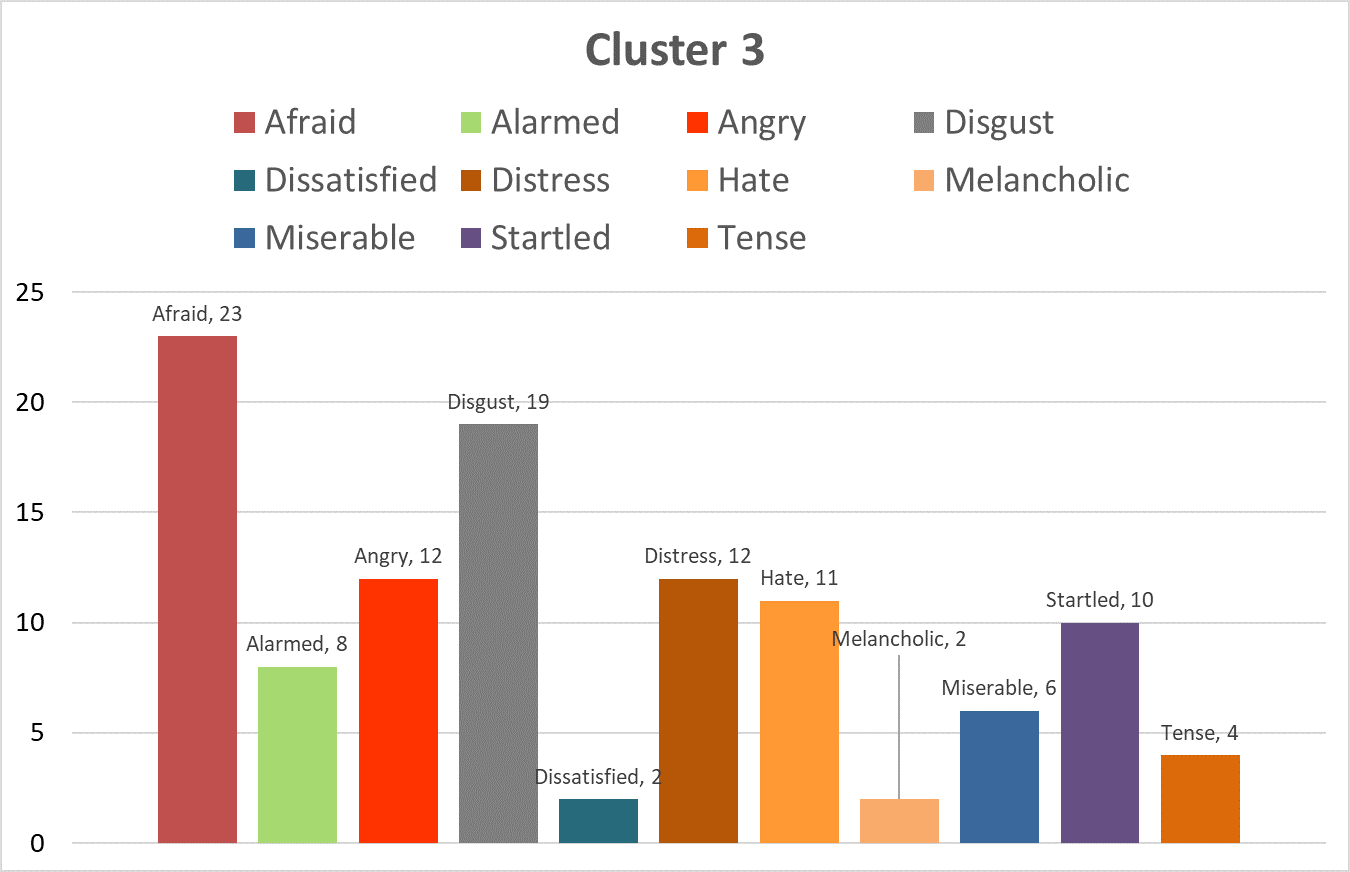}
  \end{subfigure}
\caption{Emotion Distributions in All the four FCM partitions. Many emotions are partitioned in more than two clusters, having similar mean Valence and Arousal values but significantly different mean Dominance values.}
\label{fig:cluster_emotions}
\end{figure*}

\begin{figure*}[h!]
\centering
 \begin{subfigure}{0.4\textwidth}
    \includegraphics[width=\linewidth]{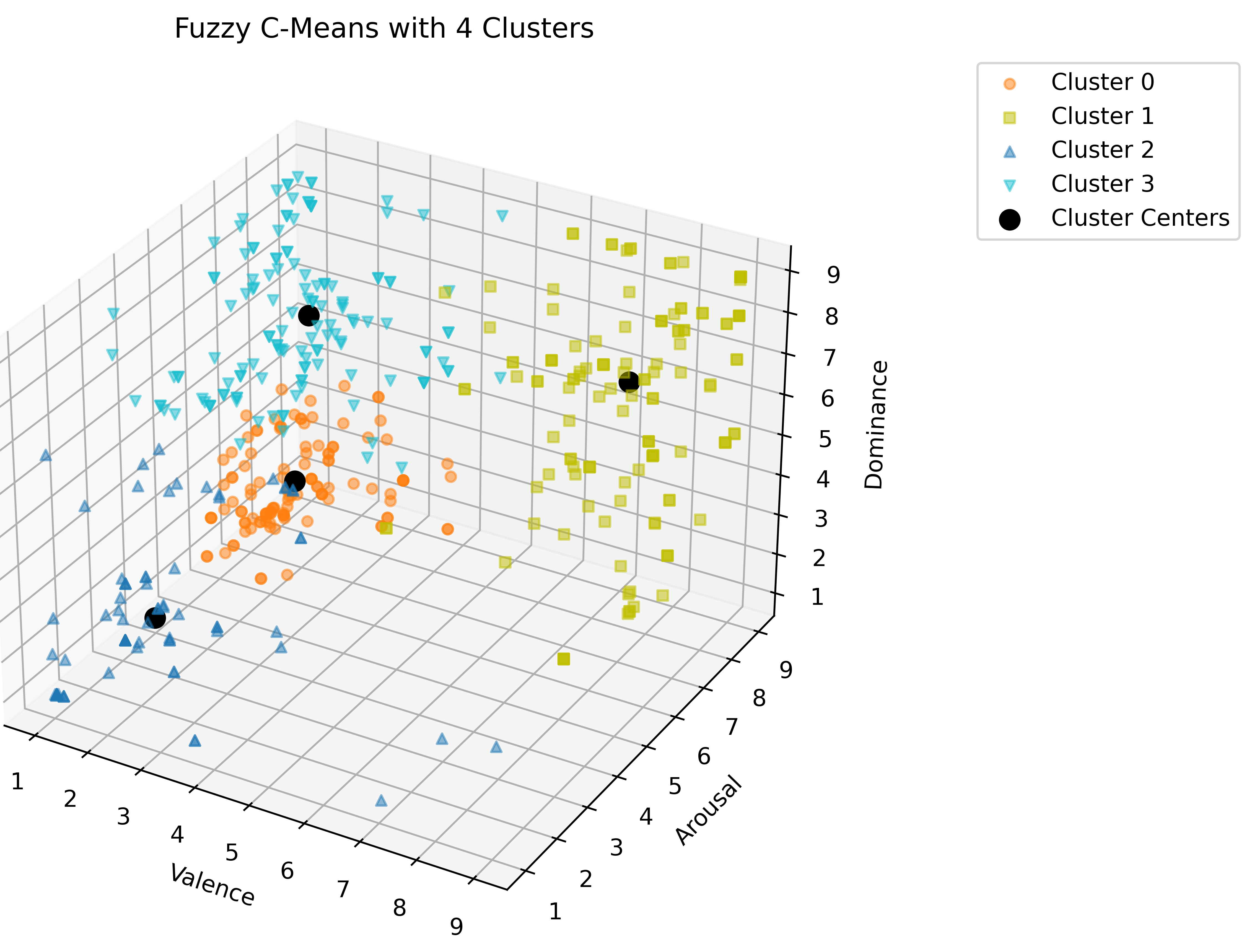}
    \caption{Fuzzy C-Means with 4 Clusters}
    \label{fig:sub1_4cluster_comparison}
  \end{subfigure}
  \hfill
  \begin{subfigure}{0.59\textwidth}
    \includegraphics[width=\linewidth]{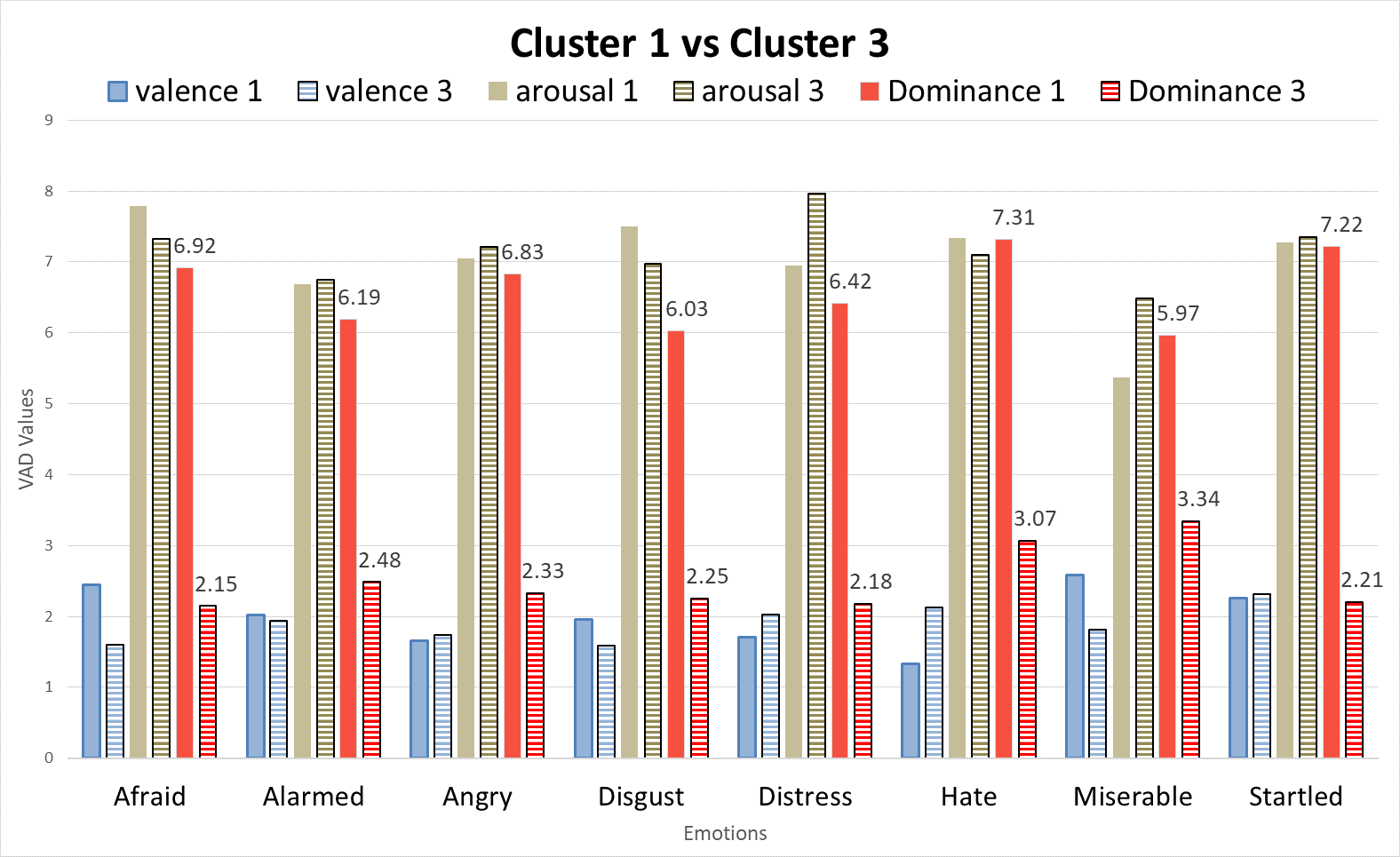}
    \caption{Emotions with same labels, but in different clusters}
    \label{fig:sub2_4cluster_comparison}
  \end{subfigure}

 \begin{subfigure}{0.4\textwidth}
    \includegraphics[width=\linewidth]{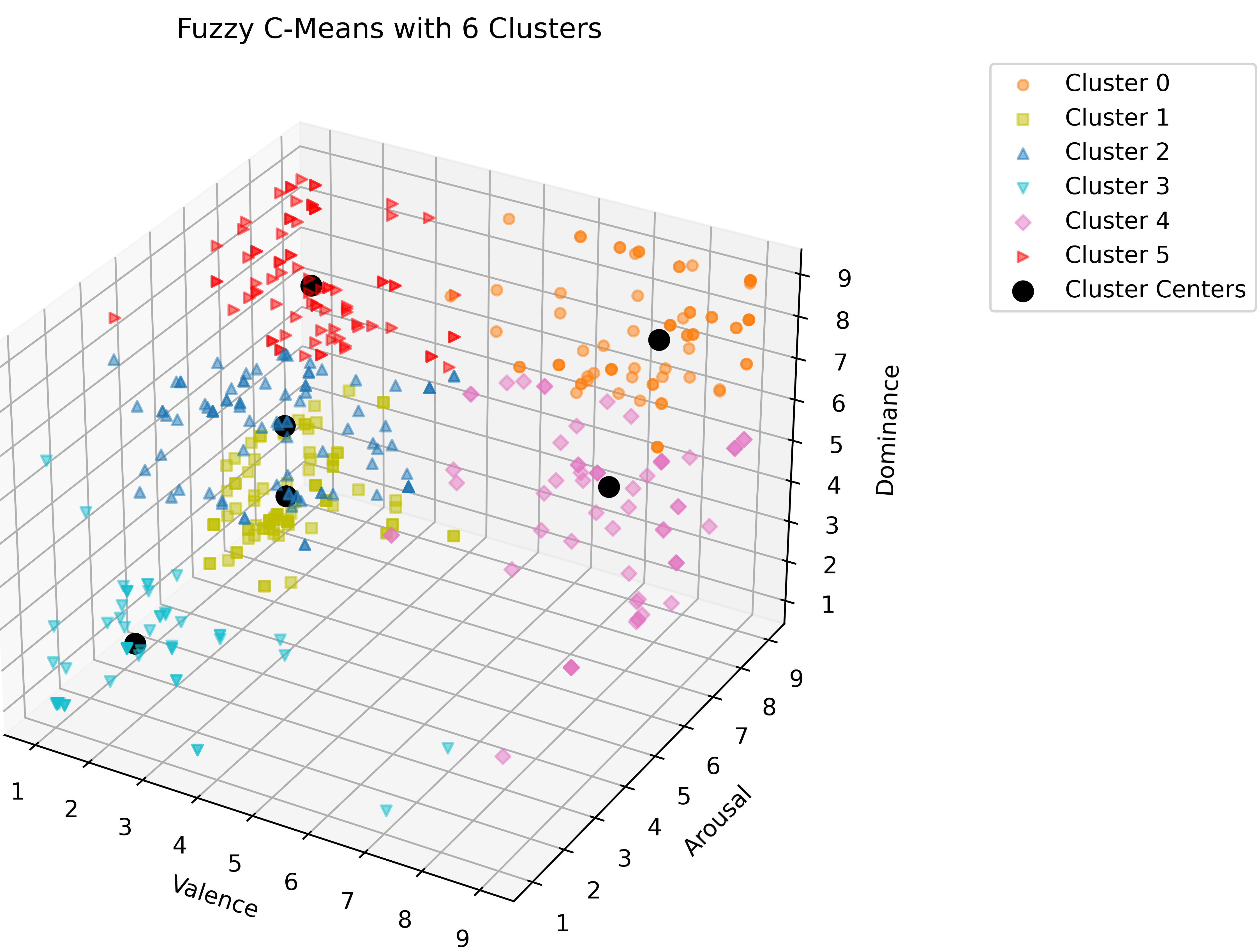}
    \caption{Fuzzy C-Means with 6 Clusters}
    \label{fig:sub1_6cluster_comparison}
  \end{subfigure}
  \hfill
  \begin{subfigure}{0.59\textwidth}
    \includegraphics[width=\linewidth]{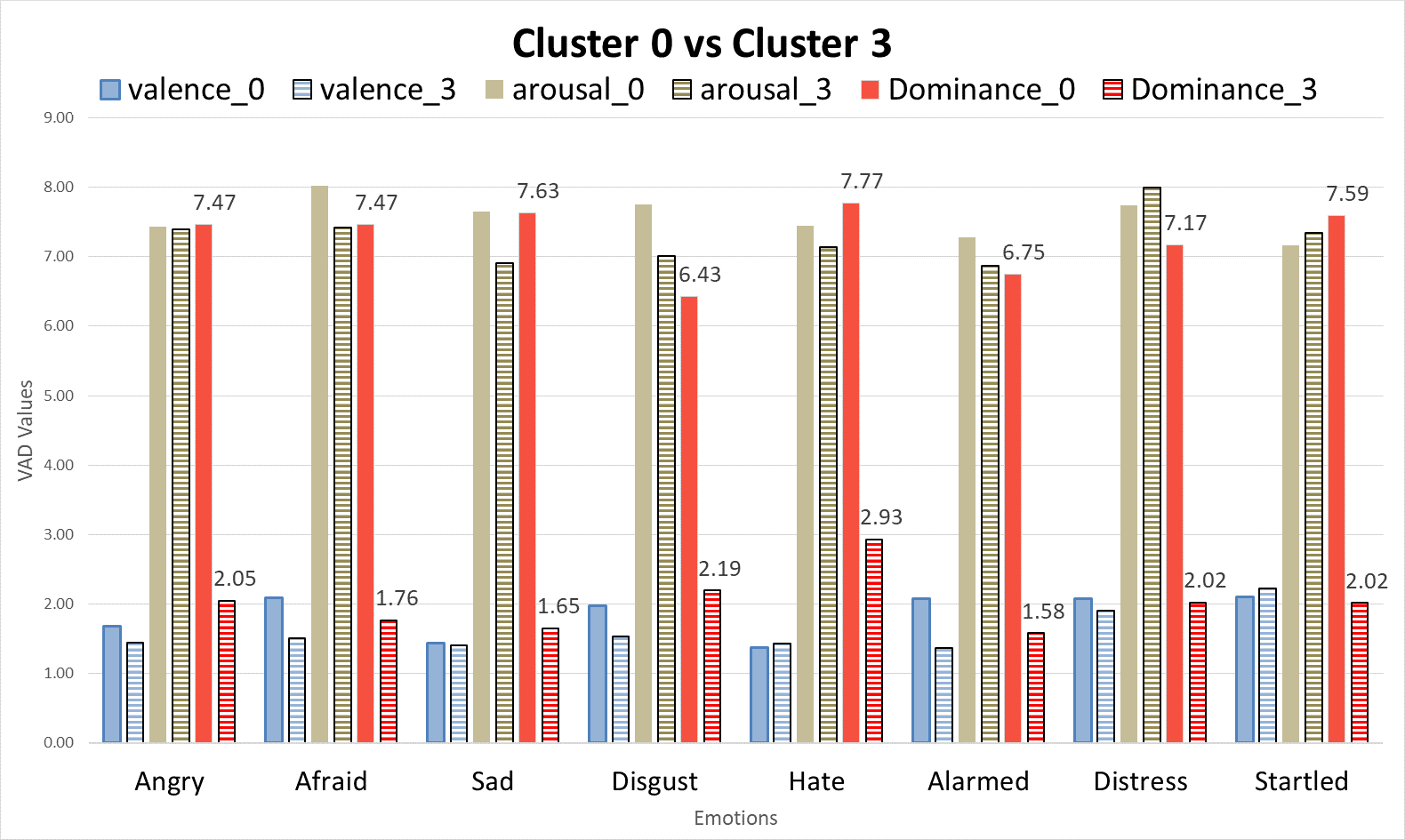}
    \caption{Emotions with same labels, but in different clusters}
    \label{fig:sub2_6cluster_comparison}
  \end{subfigure}

\caption{Fuzzy C-Means with 4 and 6 Clusters: Analysis of Same Emotion in More than One Clusters.}
\label{fig:cluster_comparison}
\end{figure*}

\section{Beyond Traditional Labels: Mapping Emotions in the VAD Space using DENS}
In this section, we give a description of the emotional dimensions and empirically analyse why considering emotions in an emotional dimension (VAD space) is crucial. We also present an overview of the emotions collected from the DENS dataset\cite{10101783}, which covers twenty-four different emotions. These emotions have concise information of the time and have EEG recordings with subjective ratings.

\subsection{Exploring the Dimensions of Emotional Experiences}
Emotions are often represented in a three-dimensional space defined by valence, arousal, and dominance, known as the VAD model \cite{mehrabian1974approach, warriner2013norms}. The representation in this space is a potential way to capture the multifaceted nature of human emotions \cite{mishra2022cardiac, mishra2022dynamic}. This section provides an in-depth overview of the VAD space dimensions, elucidating their individual significance and collective role in portraying the richness of emotional experiences. Understanding the interaction of these three dimensions allows for a subtle and comprehensive representation of emotions. The combination of valence, arousal, and dominance provides a geometric space where each point corresponds to a unique emotional state.

When it comes to understanding emotions, researchers in affective neuroscience and psychology often explore how emotions are represented in the mind and how people subjectively experience them \cite{feldman1998independence, barrett2006emotions}. Affective experiences are central to our understanding of emotions, and these experiences are not merely influenced by language or social desirability but are intrinsic to how emotions are mentally represented \cite{barrett2006emotions}. Hence, it is important to understand the subjective experiences of emotions and how they are mentally represented in order to gain a deeper understanding of how emotions influence our thoughts, behaviours, and overall well-being. While core affect, or the basic experience of pleasure or displeasure, provides a foundational aspect of emotional experience, it is acknowledged that a complete understanding of emotions goes beyond this basic level. As mentioned, an experience of emotion is not just about feeling pleasure or displeasure; it is an intentional state that involves being directed towards or about something \cite{barrett2004arousal}. 

The level of physiological arousal (e.g., heart rate, breathing, muscle tension) is a crucial component of emotional experiences. Different emotions are associated with varying levels of arousal. For example, excitement or fear may be linked to high arousal, while calmness or sadness may be associated with lower arousal \cite{kreibig2010autonomic}.

Having these two dimensions is not sufficient if we ignore the degree of control or influence one perceives they have over their emotional experience and expression. This is the Dominance dimension we refer to, and it is one of the key dimensions often used to characterise emotions, along with valence (pleasantness/unpleasantness) and arousal (intensity or activation level) \cite{kuppens2013relation}.

On one end of the spectrum, there are emotions that are more dominant or controllable. These are the emotions where individuals feel like they have agency over their emotional experiences. They can regulate their emotions consciously and choose how to respond to a situation. On the other end of the spectrum, there are emotions that are less controllable or more dominant over an individual's subjective experience. In these cases, it might feel like the emotion takes control rather than being consciously regulated\cite{jerram2014neural}.

There are individual differences in how people perceive and experience control over their emotions. Overall, recognising the dominance or control dimension adds depth to our understanding of emotions and contributes to the development of strategies for adaptive functioning.

\subsection{Overview of Twenty-Four Emotions and DENS Dataset}
The research employed naturalistic stimuli \cite{mishra2023affective} to elicit emotional responses from participants. Each participant was randomly exposed to nine emotional stimuli and two non-emotional stimuli selected from a pool of 16 emotional stimuli, each lasting 60 seconds. The dataset encompasses EEG recordings from forty participants (mean age: $23.3 \pm 1.25$, F=3) while viewing emotionally evocative films. Participants were instructed to perform a mouse click upon experiencing any emotional response, referred to as an Emotional Event. Post each video stimulus, participants assessed six dimensions using self-assessment scales measuring valence, arousal, dominance, liking, familiarity, and relevance, with valence, arousal, and dominance scales ranging from 1 to 9.

Participants selected emotions from a predefined list or specified an emotional category aligned with their experience, resulting in 24 labelled emotions. The dataset also includes 10 other emotions associated with fewer emotional events.

A total of 465 emotional experiences were collected, with each participant clicking at least once, averaging 1.29 clicks per participant. Though EEG recordings covered the entire stimulus duration (60 seconds), the study focused on a specific 7-second segment for each emotional event, yielding optimal results compared to alternative timeframes (8s, 9s, and 10s). The recordings were sampled at 250 Hz.

There are a total of 465 recorded emotional events. The distribution of valence, arousal and dominance is shown in Fig.\ref{fig:VAD_distribution} of all the 465 emotional events. The total number of labelled emotions is 24 in these 465 emotional events. The distribution of all the twenty-four emotions in VAD space is shown in Fig.\ref{fig:Distribution}. Emotion percentage distribution of these twenty-four emotions in all the emotional events is presented in Fig.\ref{fig:emotion_wheel}.

\subsection{Crucial Empirical Insights: Significance of Emotional Dimensions, with Emphasis on Dominance in VAD Space}

To gain a better empirical analysis of the relationship between multidimensional space and emotions, we used Fuzzy C-Means (FCM) to divide the emotions in different clusters. We determined the optimal number of clusters using the fuzzy silhouette index\cite{campello2006fuzzy} and found the optimum number of clusters is 4. FCM is applied to the VAD values for all the emotional events. We assigned all the emotional labels to these clusters based on the highest membership value in each cluster\ref{tab:FCM_clusters} \ref{fig:sub1_4cluster_comparison}. Emotion distribution to these clusters is shown in Fig. \ref{fig:cluster_emotions}.

\begin{table}[h!]
    \centering
    \begin{tabular}{ccccc}
        \hline
        \multirow{2}{*}{\textbf{Clusters}} & \multicolumn{3}{c}{\textbf{Centroids}} & \multirow{2}{*}{\textbf{Number of Samples}} \\
        \cline{2-4} 
        & \textbf{Valence} & \textbf{Arousal} & \textbf{Dominance} & \\
        \hline
        Cluster-0 & 7.73 & 7.70  & 6.82 & 133 \\
        Cluster-1 & 2.09 & 7.17  & 6.75 & 140\\
        Cluster-2 & 1.89 & 2.63 & 2.14 & 73\\
        Cluster-3 & 1.86 & 7.05 & 2.57 & 119 \\
        \hline
    \end{tabular}
    \caption{Centroid of Clusters Received from FCM}
    \label{tab:FCM_clusters}
\end{table}

After determining the optimal number of clusters using the fuzzy silhouette index, fuzzy C-Means is applied to the VAD values of the emotional space, resulting in their division into four clusters. The assignment of emotional labels to these clusters was based on the highest membership value in each cluster. 

The analysis revealed that emotions like- Happy, Excited and Joyous predominantly belonged to cluster-0, which notably possessed a cluster centroid for valence 7.73, distinguishing it from other clusters whose centroid valence values fell within the range of 1 to 3. A clear distinction was observed between the Sad and Angry emotions. The sad emotions, which are less related to emotions causing agitation, were categorised into cluster 2. The discernible contrast in emotional agitation between the emotions within cluster 2 and those in clusters 1 and 3 is further evident in the disparity of centroid arousal values. Specifically, cluster 2 demonstrates a value of 2.63, while clusters 1 and 3 exhibit values of 7.17 and 7.05, respectively. Clusters 1 and 3 exhibited similar angry-afraid emotions, with varying combinations depending on the dominance of specific emotions. Emotion "Afraid" is found in two different clusters with almost equal numbers of emotional events. Upon analysis, it can be observed that the mean Valence and mean Arousal values are quite similar in these two clusters for this particular emotion. However, the mean Dominance value differs significantly between the two clusters. This is a similar case with many other emotions, as shown in Fig. \ref{fig:cluster_emotions}.

Figure\ref{fig:cluster_comparison} represent the four and six FCM clusters with respective emotions that appear in more than one group. These emotions have the same emotional labels, similar valence, and similar arousal but significantly different dominance values.
In our previous study, we found that including the Dominance dimension improves accuracy when only considering the Valence and Arousal dimensions.\cite{asif2022emotion_ihci}
These differences are the results from the subjective experiences of the participants where the participant perceives emotion as other participants perceive, but the controlling mechanism of the participants differs a lot, which can only be examined with the Dominance value. We cannot explain why this emotion should be classified separately in different clusters if we don't consider the dominance value. In conclusion, it is crucial to map an emotion into VAD space instead of representing it as a discrete entity.

\section{Beyond Binary: Practical Strategies for Interpreting Human Emotion Complexity with Fuzzy Representation}\label{sec:fuzzy_representation}
Emotions are subjective and can be ambiguous. They are represented in three dimensions, but due to individual differences in perception, the crisp values of subjective feelings and reporting may not accurately reflect the true nature of the emotion. Applying fuzzification to these dimensions reflects the complexity and subjectivity of emotional experiences. The concept of fuzzification deals with uncertainty and ambiguity by allowing values to be flexible and fall within a range instead of precise points. Fuzzification can indeed be applied to the dimensions of valence, arousal, and dominance in representing the complexity and variability of emotional experiences \cite{el2000flame, ortony2022cognitive}.

Subjects feel and try to respond to their subjective feelings in terms of crisp values. Still, it will be more realistic if they represent it in continuous values such as low, medium or high and in membership functions in defining the imprecise boundaries of emotional states such as emotion belonging to a particular group or nearness to a particular group or sharing multiple groups. This approach acknowledges the fluid and subjective nature of emotional experiences and accommodates the inherent uncertainty in emotional expression\cite{li2020roles}. This concept enables the continuous grading of emotional ratings, which traditional categorical models may struggle to capture\cite{ruef2007continuous}. 

Representing emotions through fuzzy VAD introduces the concept of soft boundaries between emotional states. This acknowledges that transitions between emotions are gradual instead of abrupt, which aligns with the dynamic and continuous nature of emotional experiences\cite{ruef2007continuous, barrett2006emotions, barrett2017emotions}. This soft transition allows for a more natural and realistic representation of the fluidity between different emotional states. Furthermore, it helps in reducing misclassifications in emotion recognition systems. By capturing the uncertainty and imprecision inherent in emotional expression, fuzzy concepts mitigate the risk of rigid categorisations that may lead to misinterpretations of complex emotional states\cite{fernandez2023emotional}. This is particularly valuable in real-world scenarios where emotions are multifaceted and context-dependent\cite{barrett2006emotions}.

Instead of strictly categorising emotions as purely positive or negative, fuzzification allows for shades of emotional experiences. Emotions may fall on a continuum of valence, acknowledging that some emotions may have mixed or nuanced valence. This idea aligns with the concept that emotions exist on a spectrum rather than being neatly categorised. Using fuzzy concepts in emotion dimensions can be a significant improvement in emotion representation methodologies, as it contains the margin for personal biases. Arousal can also be fuzzified to capture the variability in intensity experienced across different emotions. For example, an emotion could be moderately arousing rather than strictly categorised as high or low arousal\cite{barrett2004arousal}. This can manage uncertainty and imprecision, enabling us to capture the subjective and context-dependent nature of human emotions in a more subtle way. Applying fuzzy concepts in emotional dimensions offers improved representational flexibility, allowing a deeper understanding of how emotions are experienced and expressed in different contexts. Fuzzifying the dominance dimension recognises that individuals may perceive varying degrees of control over their emotions \cite{siemer2007same}. Some emotions may be perceived as partially controllable, and fuzzification allows this gradation. Additionally, it provides a flexible framework that can adapt to individual differences in emotional expression and interpretation. 

We can depict the emotion-\(i\) representation in VAD space by
\begin{equation}
    E_i \sim f(x_V,x_A,x_D)
\end{equation}
where \(E\) $\in$ \{\text{Set of Emotions}\}, \(f\) represents a function, and \( x_V,x_A,x_D \) are the crisp values of valence, arousal and dominance provided by the participants. 
A more realistic and comprehensive emotion representation of Eq.\ref{eq:crispFn} depicting real-world scenarios can be represented by
\begin{equation}\label{eq:fuzzyFn}
    E_i \sim f(\mu_V(x_V),\mu_A(x_A),\mu_D(x_D))
\end{equation}
where \(E\) $\in$ \{\text{Set of Emotions}\}, \(f\) represents a function, and \( \mu_V(x_V),\mu_A(x_A),\mu_D(x_D) \) are the membership degrees of valence, arousal and dominance dimensions for their respective crisp values of \( x_V,x_A,x_D \). 
\linebreak

\noindent {\bf{Type-2 Fuzzy Representation of Emotions:}}

\begin{figure}
\centering
 \begin{subfigure}{\textwidth}
    \includegraphics[width=.49\linewidth]{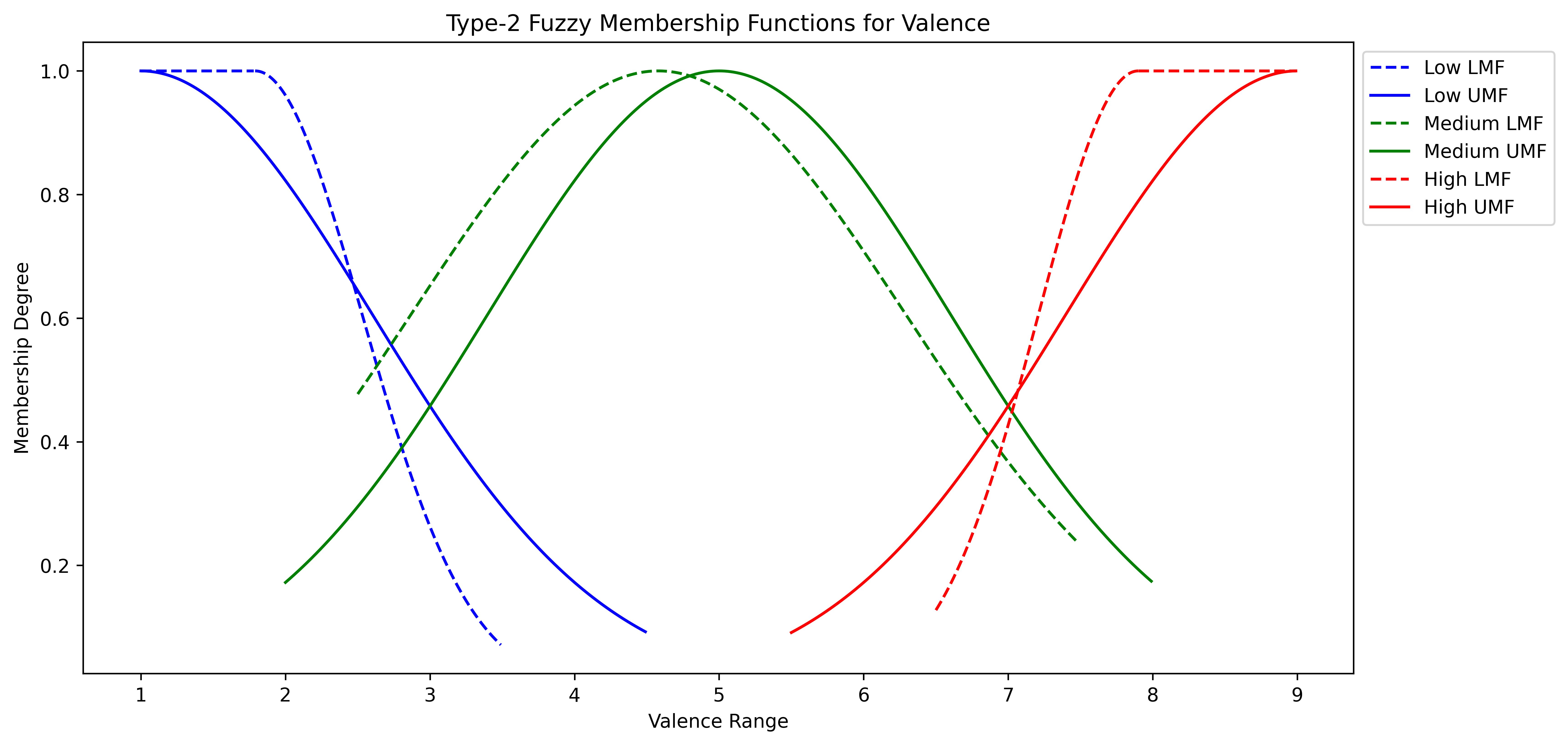}
    \label{fig:sub1_mf}
  \end{subfigure}
  \begin{subfigure}{\textwidth}
    \includegraphics[width=.49\linewidth]{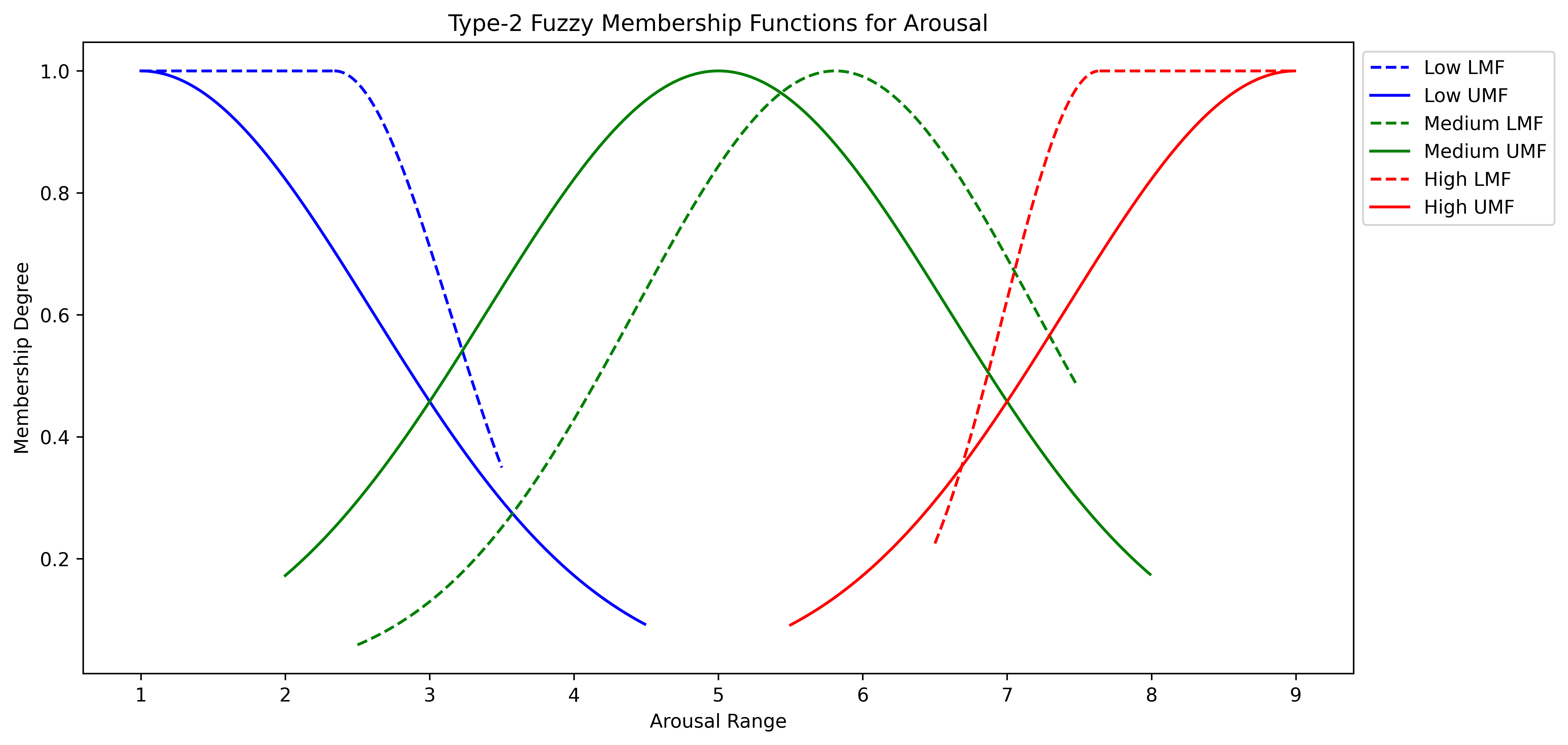}
    \label{fig:sub2_mf}
  \end{subfigure}
    \begin{subfigure}{\textwidth}
    \includegraphics[width=.49\linewidth]{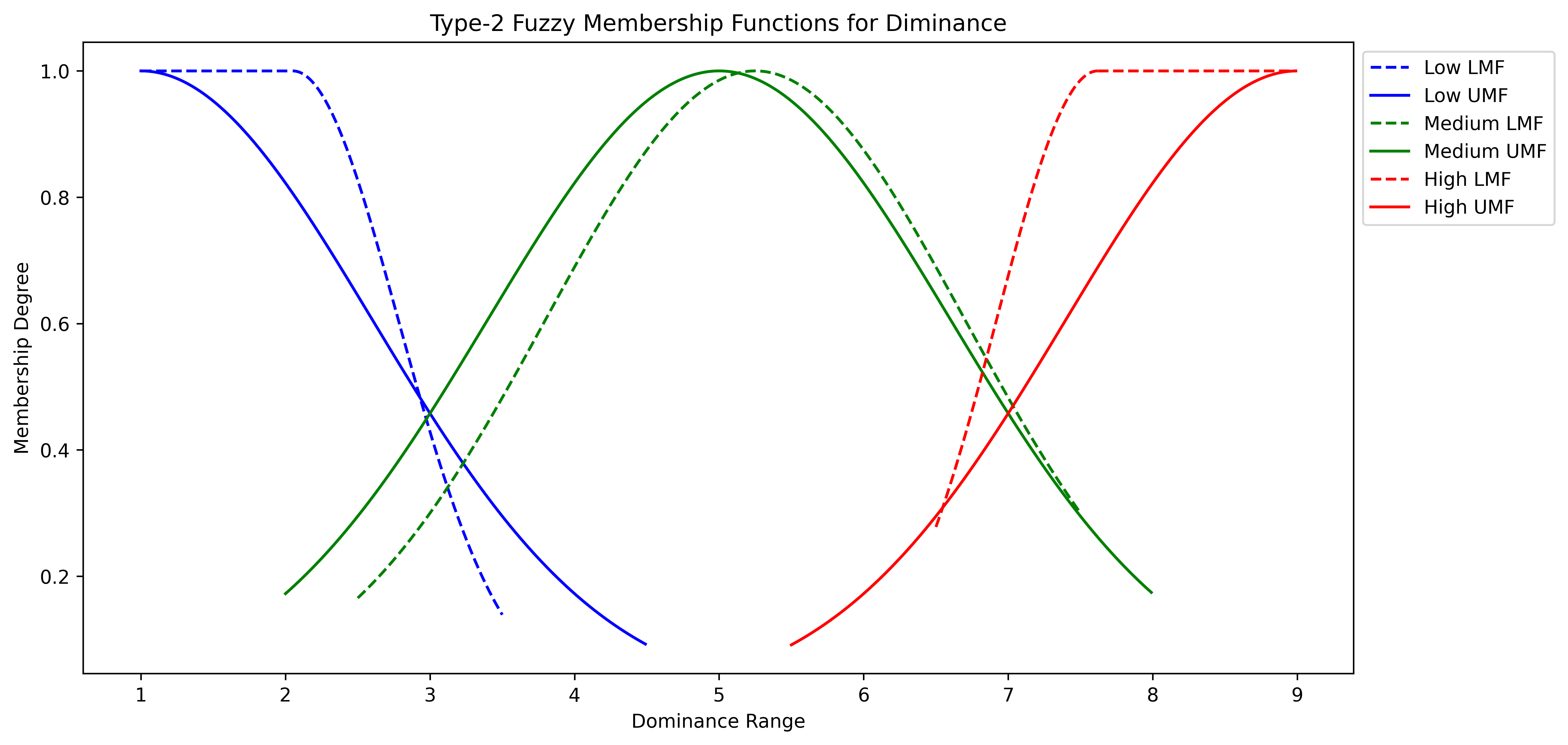}
    \label{fig:sub3_mf}
  \end{subfigure}

\caption{Membership Functions Used in VAD. Ranges are set as: $\text{Low}^\text{UMF}_\text{u} = 4.5 , \text{Med}^{\text{UMF}}_l=2, \text{Med}^{\text{UMF}}_u=8, \text{High}^{\text{UMF}}_\text{l}=5.5, \text{Low}^\text{LMF}_\text{u} = 3.5 , \text{Med}^{\text{LMF}}_l=2.5, \text{Med}^{\text{LMF}}_u=7.5, \text{High}^{\text{UMF}}_\text{l}=6.5$. Other statistical details are given in Table\ref{tab:mf_stats}}
\label{fig:mf}
\end{figure}

We defined the representation of emotion with interval type-2 fuzzy sets. The two membership functions- Lower and Upper (LMF and UMF), are based on the general meaning of the terms and the population-based meaning of the terms. UMF is defined as a generic term for all the dimensions, while LMF is set based on the statistical values of the population. In this representation, crisp values of the VAD are converted into phonetic representations of low, medium and high. Membership functions are drawn in  Fig. \ref{fig:mf}, and membership degrees are defined as follows:

Upper Membership Functions (UMF):

\begin{equation}\label{eq:mf1}
    \mu_{\text{Low}}^{\text{UMF}}(X^{\text{dim}}) = \begin{cases} 
  e^{-\frac{{(X^{\text{dim}} - \text{M}_{\text{Low}}^{\text{UMF}})^2}}{({\sigma_{\text{Low}}^{\text{UMF}})^2}}} & \text{for } \text{X}^{\text{dim}} \leq \text{Low}^\text{UMF}_\text{u} \\
  0 & \text{for } \text{X}^{\text{dim}} > \text{Low}^\text{UMF}_\text{u}
\end{cases}
\end{equation}

\begin{equation}\label{eq:mf2}
    \mu_{\text{Med}}^{\text{UMF}}(X^{\text{dim}}) = \begin{cases} 
  0 & \text{for } X^{\text{dim}} < \text{Med}^{\text{UMF}}_l \\
  e^{-\frac{{(X^{\text{dim}} - \text{M}_{\text{Med}}^{\text{UMF}})^2}}{{(\sigma_{\text{Med}}^{\text{UMF}})^2}}} & \text{for } \text{Med}^\text{UMF}_\text{l} \leq X^{\text{dim}} \leq \text{Med}^\text{UMF}_\text{u} \\
  0 & \text{for } X^{\text{dim}} > \text{Med}^\text{UMF}_\text{u}
\end{cases}
\end{equation}

\begin{equation}\label{eq:mf3}
    \mu_{\text{High}}^{\text{UMF}}(X^{\text{dim}}) = \begin{cases} 
  0 & \text{for } X^{\text{dim}} < \text{High}^{\text{UMF}}_l \\
  e^{-\frac{{(X^{\text{dim}} - \text{M}_{\text{High}}^{\text{UMF}})^2}}{{(\sigma_{\text{High}}^{\text{UMF}})^2}}} & \text{for } X^{\text{dim}} \geq \text{High}^{\text{UMF}}_\text{l} \\
\end{cases}
\end{equation}

Lower Membership Functions (LMF):
\begin{equation}\label{eq:mf4}
\mu_{\text{Low}}^{\text{LMF}}(X^{\text{dim}}) = \begin{cases} 
  1 & \text{for } \text{X}^{\text{dim}} \leq \text{M}_{\text{Low}}^{\text{LMF}} \\
  e^{-\frac{{(X^{\text{dim}} - \text{M}_{\text{Low}}^{\text{LMF}})^2}}{({\sigma_{\text{Low}}^{\text{LMF}})^2}}} & \text{for } \text{M}_{\text{l}}^{\text{LMF}} < \text{X}^{\text{dim}} \leq \text{Low}^\text{LMF}_\text{u} \\
  0 & \text{for } \text{X}^{\text{dim}} > \text{Low}^\text{UMF}_\text{u}
\end{cases}
\end{equation}

\begin{equation}\label{eq:mf5}
    \mu_{\text{Med}}^{\text{LMF}}(X^{\text{dim}}) = \begin{cases} 
  0 & \text{for } X^{\text{dim}} < \text{Med}^{\text{LMF}}_l \\
  e^{-\frac{{(X^{\text{dim}} - \text{M}_{\text{Med}}^{\text{LMF}})^2}}{{(\sigma_{\text{Med}}^{\text{LMF}})^2}}} & \text{for } \text{Med}^\text{LMF}_\text{l} \leq X^{\text{dim}} \leq \text{Med}^\text{LMF}_\text{u} \\
  0 & \text{for } X^{\text{dim}} > \text{Med}^\text{LMF}_\text{u}
\end{cases}
\end{equation}

\begin{equation}\label{eq:mf6}
    \mu_{\text{High}}^{\text{LMF}}(X^{\text{dim}}) = \begin{cases} 
  0 & \text{for } X^{\text{dim}} < \text{High}^{\text{LMF}}_l \\
  e^{-\frac{{(X^{\text{dim}} - \text{M}_{\text{High}}^{\text{LMF}})^2}}{{(\sigma_{\text{High}}^{\text{LMF}})^2}}} & \text{for } \text{High}^{\text{LMF}}_\text{l} \leq X^{\text{dim}} > \text{M}_{\text{High}}^{\text{LMF}} \\
  1 & \text{for } X^{\text{dim}} \geq \text{M}_{\text{High}}^{\text{LMF}} \\
\end{cases}
\end{equation}

The membership function, denoted by \( \mu \), represents the degree of membership. 
\(dim\) represents the emotion dimension i.e., $ dim \in \{V, A, D\}$.

The symbol \(X^{\text{dim}}\) represents the input value for that dimension and \(X^{\text{dim}}\) $\in [1, 9]$.

$\text{M}$ and $\sigma$ represent the mean and standard deviation of the various ranges for Gaussian distributions.

$Range(\text{Low}^{\text{UMF}}) \in [1,\text{Low}^\text{UMF}_\text{u}]$

$Range(\text{Med}^{\text{UMF}}) \in [\text{Med}^\text{UMF}_\text{l}, \text{Med}^\text{UMF}_\text{u}]$

$Range(\text{High}^{\text{UMF}}) \in [\text{High}^\text{UMF}_\text{l},9]$

$Range(\text{Low}^{\text{LMF}}) \in [1,\text{Low}^\text{LMF}_\text{u}]$

$Range(\text{Med}^{\text{LMF}}) \in [\text{Med}^\text{LMF}_\text{l}, \text{Med}^\text{UMF}_\text{u}]$

$Range(\text{High}^{\text{LMF}}) \in [\text{High}^\text{LMF}_\text{l},9]$

The membership functions are shown in Fig.\ref{fig:mf}. The corresponding mean, standard deviation and ranges are shown in Table\ref{tab:mf_stats}. These membership functions are modified where they intersect to define the Footprint of Uncertainty (FoU), and the means are adjusted accordingly. 
\begin{equation}\label{eq:FoUFn}
    FoU = F(\mu^\text{UMF}, \mu^\text{LMF})
\end{equation}
where F represents a function to determine the FoU and coincide the means to a single point. $\mu^\text{UMF}$ and  $\mu^\text{LMF}$ represent the UMF and LMF memberships.

\begin{table*}
    \centering
    \begin{tabular}{c|cc|cc|cc||cc|cc|cc}
    \hline
        Dimension & 
        $\text{M}_{\text{Low}}^{\text{UMF}}$ & $\sigma_{\text{Low}}^{\text{UMF}}$ & 
        $\text{M}_{\text{Med}}^{\text{UMF}}$ & $\sigma_{\text{Med}}^{\text{UMF}}$ & 
        $\text{M}_{\text{High}}^{\text{UMF}}$ & $\sigma_{\text{High}}^{\text{UMF}}$ & 
        $\text{M}_{\text{Low}}^{\text{LMF}}$ & $\sigma_{\text{Low}}^{\text{LMF}}$ & 
        $\text{M}_{\text{Med}}^{\text{LMF}}$ & $\sigma_{\text{Med}}^{\text{LMF}}$ & 
        $\text{M}_{\text{High}}^{\text{LMF}}$ & $\sigma_{\text{High}}^{\text{LMF}}$ \\
        \hline
        Valence & 1 & 1.2 & 5 & 1.2 & 9 & 1.2 & 1.96 & 0.92 & 4.30 & 1.97 & 7.63 & 0.92\\
        
        Arousal & 1 & 1.2 & 5 & 1.2 & 9 & 1.2 &2.8 & 1.02 & 6.04 & 1.54 & 7.26 & 0.90  \\
        
        Dominance & 1 & 1.2 & 5 & 1.2 & 9 & 1.2 & 2.40 & 1.01 & 5.02 & 1.81 & 7.27 & 0.90  \\
        \hline
    \end{tabular}
    \caption{Means and Standard Deviations for Various Populations. LMF means and standard deviations are based on real population, while UMF means and standard deviations are given based on the generic understanding of the perception.}
    \label{tab:mf_stats}
\end{table*}

\section{Methodology}

\subsection{Data Preparation}

\subsubsection{Emotional Events Extraction and Preprocessing}
In the initial phase of EEG signal processing, the raw data, initially referenced to the Cz electrode, was imported. Subsequently, average re-referencing was applied. The EEG signal underwent filtering using a fifth-order Butterworth bandpass filter, with a low cutoff frequency of 1 Hz and a high cutoff frequency of 40 Hz. Segmentation of the filtered signal was performed to isolate event segments corresponding to the baseline state and specific emotional events (captured using a mouse clicks). The baseline state was defined as the period from 10 to 70 seconds, while the click events were extracted from -6 to +1 seconds relative to the click response for the emotional event by the participants. The duration of the extracted click events spanned from 6 seconds before to 1 second after the event.

Following signal concatenation of extracted emotional events, a manual inspection was conducted to ensure data quality. Electrodes and samples displaying high amplitude, potentially due to electrode separation, were excluded from the analysis.

To mitigate the impact of eyeblink activity, Independent Component Analysis (ICA) was employed. This step was crucial to prevent automatic channel rejection based on high amplitudes associated with eye artefacts. After the initial automatic channel rejection, a second ICA iteration was implemented to eliminate any residual artefacts, such as muscle and heart activity, line noise, and channel noise. The ICLabel tool categorized independent components into different labels\cite{pion2019iclabel}, and components with a brain activity probability greater than 0.3 were retained. The probabilities for all labels equal to 1 were aggregated.

Upon completion of the pre-processing steps, discrete occurrences corresponding to the baseline state and emotional responses were stored, facilitating subsequent data analysis.

\subsubsection{Feature Extraction and Input Formatting}
The Short-Time Fourier Transform (STFT) is a widely used signal processing technique, particularly for analyzing time-varying signals like EEG data. It enables the analysis of a signal's frequency content as it evolves over time, which is critical for understanding dynamic processes such as changes in brain activity captured by EEG signals. Since EEG signals are non-stationary, meaning that their statistical properties change over time, STFT provides a local analysis that adapts to these changes, making it well-suited for analyzing signals with varying characteristics. STFT offers the ability to capture temporal and frequency information simultaneously. Its adaptability to non-stationary signals makes it particularly well-suited for studying dynamic brain activities.

The data received from the $i^\text{th}$ EEG channel is $x_i$, The STFT of $x_i$ is represented as
\[
X_i[m, n] = \sum_{k=0}^{N-1} x_i[n + k] \cdot w[k - m] \cdot e^{-j2\pi fk/N}
\]
where 
STFT coefficient \(X_i\) represented at a specific time index \(n\) and frequency index \(m\) for the signal \(x_i[n]\). A sliding window approach is used, where the input signal \(x_i\) is shifted by \(k \) samples. The term \(w[k - m]\) denotes the value of the window function at the relative position \(k - m\). The application of the window function is integral in minimizing spectral leakage and refining the precision of the STFT. The complex exponential modulation term \(e^{-j2\pi fk/N}\) introduces a frequency-dependent weighting factor. Here, \(f\) denotes the frequency index, \(k\) represents the sample index within the window, and \(N\) stands for the total number of samples. This modulation is pivotal in elucidating the frequency content within the defined window. We used 50\% overlapping to compute the STFT.

The spectrograms are computed by
\[
S_i[m, n] = \left| X_i[m, n] \right|^{2}
\]
i.e.,
\begin{equation}
S_i[m, n] = \left| \sum_{k=0}^{N-1} x_i[n + k] \cdot w[k - m] \cdot e^{-j2\pi fk/N} \right|^{2}
\end{equation}

After this, spectrograms are stacked further to create an input to be given in the architecture as Input1, \[S_{l,m,n} \coloneqq S_i[l,m, n]\] where \(l\) is the number of stacked spectrograms.

\subsection{Architecture with Deep Fuzzy Framework}

Rather than being distinct groups, emotions can be seen as continuous values in the VAD space. This suggests that affective states and emotions are interconnected and interdependent and connected to each other systematically. Our framework is based on this phenomenon.

\begin{figure*}
\centering
\includegraphics[width=\linewidth]{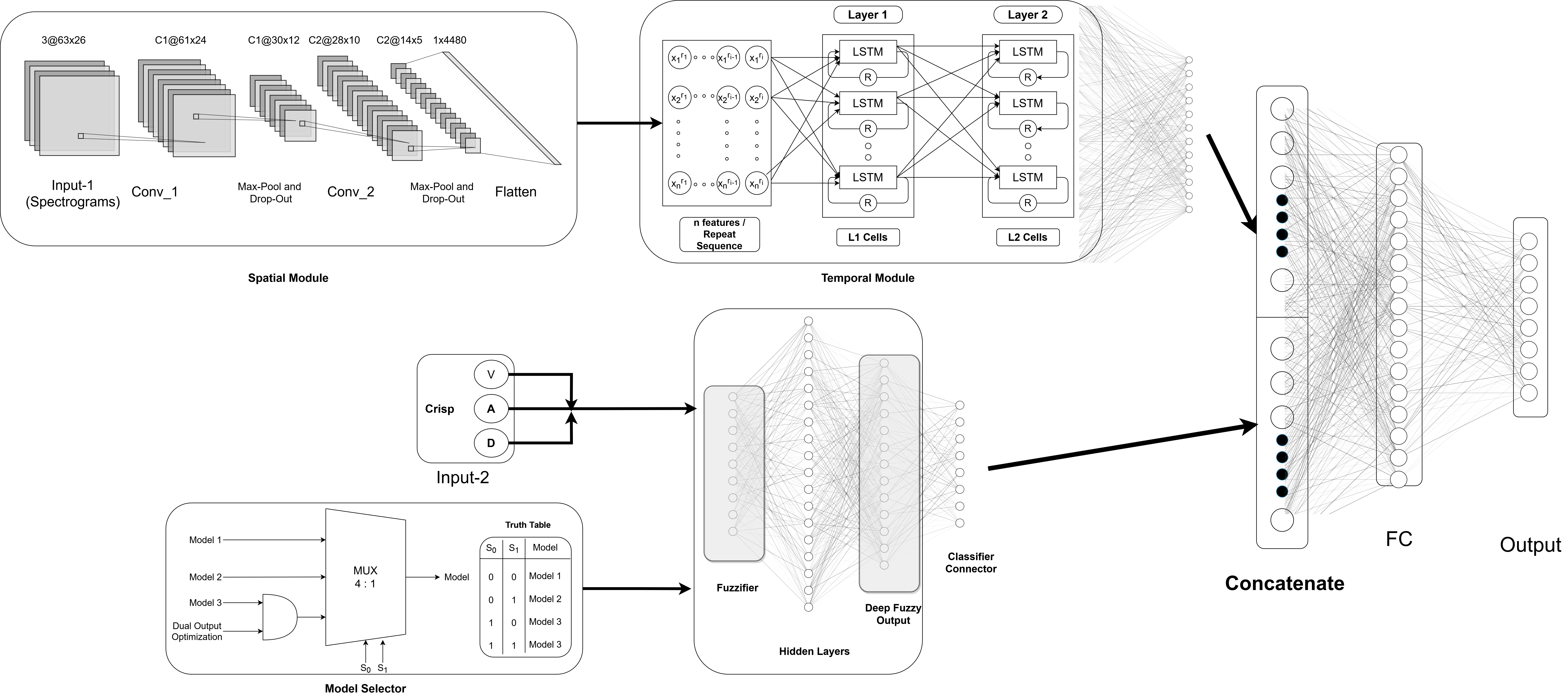}
\caption{Deep Fuzzy Architecture: This architecture consists of three modules. The spatial module is based on CNNs, the Temporal module is based on LSTMs, and the third module is our proposed Fuzzy Framework. This framework further represents three more models based on VAD fuzzy representations.}
\label{fig:architecture}
\end{figure*}

In Fig.\ref{fig:architecture}, we have presented the overall framework of our architecture, consisting of three modules. The first module captures spatial features, while the second module captures temporal features. The third module utilizes a fuzzy framework, demonstrated in Eq. \ref{eq:fuzzyFn}, to provide the membership degree of emotion to the VAD space. We built three distinct models based on this framework that contribute to three different representations of emotions, which are discussed in detail in this section.

\subsubsection{Role of Fuzzy VAD Representation in Enhancing Emotion Recognition}
The integration of VAD space with multimodal inputs, such as physiological signals (e.g., EEG), further enhances the capabilities of recognition systems. This integration allows for a more holistic understanding of emotional states by leveraging a diverse set of features. VAD space mitigates ambiguity, a common challenge in emotion recognition. The multidimensional information of emotional experience provided by VAD dimensions reduces the likelihood of misinterpretations, enhancing the robustness of emotion recognition algorithms. VAD space might also contribute to improved personalisation in emotion recognition. Recognising that individuals may exhibit varied emotional responses to the same stimuli, VAD dimensions allow recognition systems to account for these individual differences, leading to more accurate and personalised emotion predictions. The consideration of valence, arousal, and dominance provides richer information for understanding emotional responses and empowering emotion recognition systems. 

\subsubsection{Modules in the Architecture}
We have three modules in the architecture: First for spatial features, second for temporal features, and third for fuzzy representational block. Our emotion recognition architecture leverages both spatial and temporal information of EEG signals. To extract significant patterns and features from spectrograms, we used Convolutional Neural Networks (CNNs). Moreover, Long Short-Term Memory (LSTM) networks are highly effective in analyzing time-series data and modelling temporal dependencies. As a result, they are an ideal solution for this emotion recognition from EEG. Lastly, we have a parallel module for providing a VAD representational block.

\paragraph{Spatial Module}
In this process, we use kernels in the convolution process to learn local patterns from the input data. These patterns are then combined and refined through multiple layers of CNNs to yield more complex and meaningful features. We utilized two layers of CNN in our models. All the layers have dropout masks with a 20\% dropout rate. The output of the spatial module is calculated by the following set of equations:
\begin{equation}
\begin{aligned}
Y_{i,j,k}^{[1]} &= \text{MaxPooling}(\text{ReLU}(\text{Conv}(X, W^{[1]}, b^{[1]})), p_m, p_n) \\
Y_{i,j,k}^{[2]} &= \text{MaxPooling}(\text{ReLU}(\text{Conv}(Y^{[1]}, W^{[2]}, b^{[2]})), p_m, p_n) \\
Y_{\text{flattened}}^{[2]} &= \text{Flatten}(Y^{[2]}) \\
\end{aligned}
\end{equation}\label{cnn}
where 

$Y_{i,j,k}^{[1]}$ and $Y_{i,j,k}^{[2]}$ represent the outputs of the first and second convolutional layers, respectively at position (i,j,k). $X$ is the Input tensor. $W^{[1]}, W^{[2]}$ represent weights of the convolutional layers. $b^{[1]}, b^{[2]}$ are the biases of the convolutional layers. $Y_{\text{flattened}}^{[2]}$ represents the flattened output tensor from the second convolutional layer. $(p_m, p_n)$ is the window size of the pooling layer. All the layers have dropout masks with a 20\% dropout rate.

The convolution operation, denoted as $\text{Conv}(X, W^{[1]}, b^{[1]})$, is applied to the input tensor $X$ using weights $W^{[1]}$ and biases $b^{[1]}$. Following the convolution, the Rectified Linear Unit (ReLU) activation function, $\text{ReLU}(\cdot)$, is applied element-wise to the result. Mathematically, ReLU is defined as $\text{ReLU}(x) = \max$. Finally, a max-pooling operation, denoted as $\text{MaxPooling}(\cdot, p_m, p_n)$, is performed on the output with a window size of $p_m \times p_n$. We used a $(2\times2)$ pooling channel with a $(3\times3)$ kernel size of the convolution. We used 32 and 64 kernels for first and second CNN layers, respectively.

$\text{Conv}$ operation we used as-
\begin{equation}
Y_{i,j,k} = \sum_{a=1}^{k_m} \sum_{b=1}^{k_n} \sum_{c=1}^{l} W_{a,b,c,k} \cdot X_{i+a-1,j+b-1,c} + b_k
\end{equation}

The symbol $Y_{i,j,k}$ denotes the element at position $(i, j, k)$ in the output tensor. Similarly, $W_{a,b,c,k}$ represents the weight located at position $(a, b, c, k)$ within the filter $W$. Moving to the input tensor, $X_{i+a-1,j+b-1,c}$ corresponds to the element situated at position $(i+a-1, j+b-1, c)$. Additionally, $b_k$ is the bias term associated with the $k$-th filter of size $(k_m \times k_n)$. $Y_{\text{flattened}}$ will be used for temporal module.

\paragraph{Temporal Module}
 We use a hybrid architecture for emotion categorisation to take advantage of both Convolutional Neural Networks (CNN) and Long Short-Term Memory (LSTM). The hybrid CNN-LSTM model combines the CNN's ability to extract features from data with the LSTM's ability to capture both long-term and short-term dependencies.

The output from the Flatten layer, denoted as \(Y_{\text{flattened}}^{[2]}\), is sequentially processed through two Long Short-Term Memory (LSTM) layers. Before the LSTM layers, a Repeat Sequence operation is applied. This operation, often used in sequence-to-sequence tasks, repeats the output sequence from the Flatten Vector and prepares it for further processing. $R$ represents the number of Repeat Sequence. We used R=4 for further processing of the Output received from the spatial module.

\begin{equation}
\begin{aligned}
Y_{\text{rep}} &= \text{RepeatSequence}(Y_{\text{flattened}}^{[2]}, R) \\
H_{r}^{[1]}, C_{r}^{[1]} &= \text{LSTM}(Y_{\text{rep}}, H_{r-1}^{[1]}, C_{r-1}^{[1]}) \\
H_{r}^{[2]}, C_{r}^{[2]} &= \text{LSTM}(H_{r}^{[1]}, H_{r-1}^{[2]}, C_{r-1}^{[2]})
\end{aligned}
\end{equation}

In the equations, $Y_{\text{rep}}$ Repeat Sequence operation. $H_{r}^{[1]}, C_{r}^{[1]}$ represent the hidden state and cell state of the first LSTM layer at repetition $r$, while $H_{r}^{[2]}, C_{r}^{[2]}$ denote the corresponding states for the second LSTM layer. The LSTM operations involve gating mechanisms, such as input gates, forget gates, and output gates, influencing the flow of information and memory retention in the cell states. The first LSTM layer processes the repeated sequence, and its output, along with the previous hidden and cell states, becomes the input for the second LSTM layer.

In the provided equations, the final output of the second LSTM layer is denoted as $H_{r}^{[2]}$, representing the hidden state at the final time step $r$. This hidden state encapsulates the learned representation of the input sequence after processing through the first and second LSTM layers. $H_{r}^{[2]}$ will be used as a feature representation and serves as the input for subsequent layers, which concatenates with the input from the deep fuzzy module. It captures the hierarchical dependencies and patterns in the sequential data, providing a meaningful representation for further analysis or decision-making.

\subsubsection{Integration of Deep Fuzzy Module}

\begin{figure*}
\centering
\includegraphics[width=\linewidth]{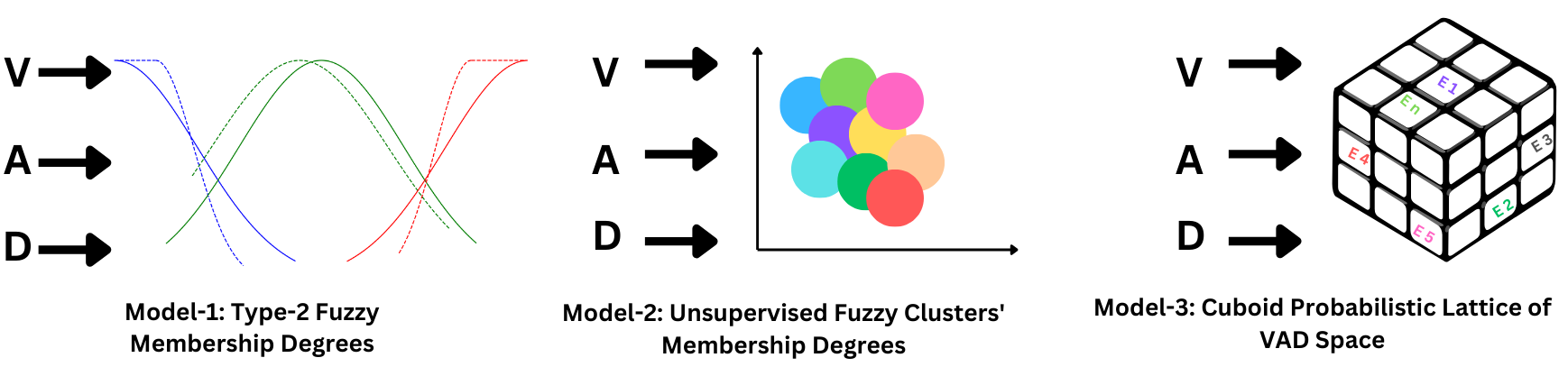}
\caption{A Schematic Representation of Developed Models: Models Based on Various Fuzzy Representations: The first model is based on the type-2 membership functions received by the UMF and LMF based on idealistic and realistic values, respectively. The next model is based on unsupervised learning and uses fuzzy C-Means membership degrees for various numbers of clusters. The last model is based on the cuboid probabilistic lattice representation of emotion in VAD space. Model-3 takes fuzzy input and optimises based on the probabilities of low, medium, and high VAD cuboids and uses dual output- one for VAD and the other for the final 24 emotions classification.}
\label{fig:models}
\end{figure*}

In our approach, we developed three models based on fuzzy representations of emotions. The aim is to use a computational approach for decision-making in the final classification. To achieve this, we utilized membership functions, unsupervised clustering, and supervised cubical probabilities. We developed three different models based on three distinct representations of emotion: firstly, degree of membership values (Type-1 and Type-2), secondly unsupervised fuzzy clusters, and thirdly, cuboid probabilistic lattice representation of the VAD dimensions. The models are schematically represented in Fig. \ref{fig:models}. The fuzzy module consists of a Fuzzifier sub-module, which converts the crisp VAD values into fuzzy membership degrees, a Model Selector, according to which membership degrees are assigned (more details about the membership degrees in individual models). and Fully Connected neural network. The last layer acts as an output for this module.
The membership degree of the \text{VAD\_Input} is,
\begin{equation}\label{eq:MF_module3}
    \Gamma = \phi(VAD\_Input)
\end{equation}
where $\phi$ is a function Fuzzifier. This function is defined differently for each model, which is discussed later in this section.

$\Gamma$ is further processed through Dense layers with ReLU and Dropout layers inbuilt with ($\cdot$) product in initial layers, while the final layer is processed differently with different models.
\[
\text{O}_{\text{FC}}^\text{[i]} = \Gamma \cdot W_{\text{FC}}^\text{[i]} + b_{\text{FC}}^\text{[i]}
\]
\begin{equation}\label{eq:module3_out}
    \text{O}_{\text{FC}}^\text{[f]} = \text{O}_{\text{FC}}^\text{[i]} \cdot W_{\text{FC}}^\text{[f]} + b_{\text{FC}}^\text{[f]}
\end{equation}

where
$\text{O}_{\text{FC}}^\text{[i]}$ and $\text{O}_{\text{FC}}^\text{[f]}$ are the outputs of the initial layers and final layers, respectively and W and b are the weight and bias, respectively.

\paragraph{Model-1}\label{sec:model1}
Model-1 uses the type-2 fuzzy membership function for $\phi$ in Eq.\ref{eq:MF_module3}. These membership functions are mentioned in Section-\ref{sec:fuzzy_representation} with the equations from Eq.\ref{eq:mf1} to Eq.\ref{eq:mf6}.

\paragraph{Model-2}\label{sec:model2}
Model-2 is based on unsupervised clustering. We used the FCM algorithm to define the degrees of each clusters in the VAD space. Here $\phi$ in Eq.\ref{eq:MF_module3} is updated as,
\begin{equation}\label{eq:FCM_model2}
    u_{ij} = \left( \sum_{k=1}^{c} \left( \frac{\lVert x_i - v_j \rVert}{\lVert x_i - v_k \rVert} \right)^{\frac{2}{m-1}} \right)^{-1}
\end{equation}
where
The membership degrees are represented by a matrix \(U\) where \(u_{ij}\) denotes the degree of membership of data point \(i\) to cluster \(j\). Here, \(i\) ranges from 1 to \(n\) (number of data points), and \(j\) ranges from 1 to \(c\) (number of clusters).

\paragraph{Model-3}\label{sec:model3}
Model-3 uses the type-2 fuzzy membership function for $\phi$ in Eq.\ref{eq:MF_module3} with the equations from Eq.\ref{eq:mf1} to Eq.\ref{eq:mf6}. However, the output layer of the fuzzy framework module in this model is used as a dual output and optimized simultaneously with the main output layer. The second output is trained on the classification of VAD space in 27 classes, just like a cuboid is divided into low, medium, and high values of its dimension. We used Softmax in the last layer of this fuzzy module, i.e., Softmax is applied on $\text{O}_{\text{FC}}^\text{[f]}$ in Eq.\ref{eq:module3_out}.
\begin{equation}\label{eq:dual_module3}
    p_i = \text{Softmax}(\text{O}_{\text{FC}}^\text{[f]})_i = \frac{e^{o_i}}{\sum_{j=1}^{k} e^{o_j}}
\end{equation}
where
i represents the index of the element in the vector, and k is the number of dense units used in the final output of the layer, i.e., the number of VAD space classes used in the second output layer. And $\mathbf{\text{O}_{\text{FC}}^\text{[f]}} = [o_1, o_2, \ldots, o_k]$. We took k=27.

We call the vector $[p_1, p_2, \ldots, p_k]$ as cuboid probabilistic lattices as it contains the probability of the VAD space mapped to k cuboids. These lattices are the output of Model-3.

\subsection{Cross-Subject Study}
In order to validate our model , we extended our work to test cross-subject emotion recognition. The cross-subject task is more complex in nature\cite{10329776}. Therefore, we simplified the task by grouping together a few emotions based on a tree-structured list\cite{shaver1987emotion, parrott2000emotions}. It is worth noticing here that we did not use VAD information to group these emotions to avoid creating any biases in the model. Three distinct groups are formed. Emotions considered in these three groups are mentioned in Table\ref{tab:cross_sub_groups}.
\begin{table}
    \centering
    \begin{tabular}{cc|cc|cc}
        \hline
        \multicolumn{2}{c|}{\textbf{Group 1}} & \multicolumn{2}{c|}{\textbf{Group 2}} & \multicolumn{2}{c}{\textbf{Group 3}} \\
        
        \textbf{Emotions} & \textbf{\#} & \textbf{Emotion} & \textbf{\#} & \textbf{Emotions} & \textbf{\#} \\
        \hline
        Delighted & 11 & Melancholic & 5 & Taken Aback & 11 \\
        Amused & 14 & Depressed & 6 & Startled & 20 \\
        Happy & 5 & Despondent & 10 & Distress & 22 \\
        Adventurous & 20 & Dissatisfied & 10 & Alarmed & 27 \\
        Joyous & 20 & Miserable & 19 & Afraid & 46 \\
        Excited & 26 & Sad & 45 &  &  \\
        \hline
        Total & 108 & Total & 95 & Total & 126 \\
        \hline
    \end{tabular}
    \caption{Group Formation for Cross-Subject Emotion Recognition}
    \label{tab:cross_sub_groups}
\end{table}

We split the samples in such a way that the training and validation sets do not have data from the same participants.

\section{Experiments and Results}
We have created three models, as discussed earlier. We performed the experiments with all the models to verify the sanctity of our framework and methodology. 

\begin{figure*}
\centering
 \begin{subfigure}{0.49\textwidth}
    \includegraphics[height=0.6\textwidth, width=\linewidth]{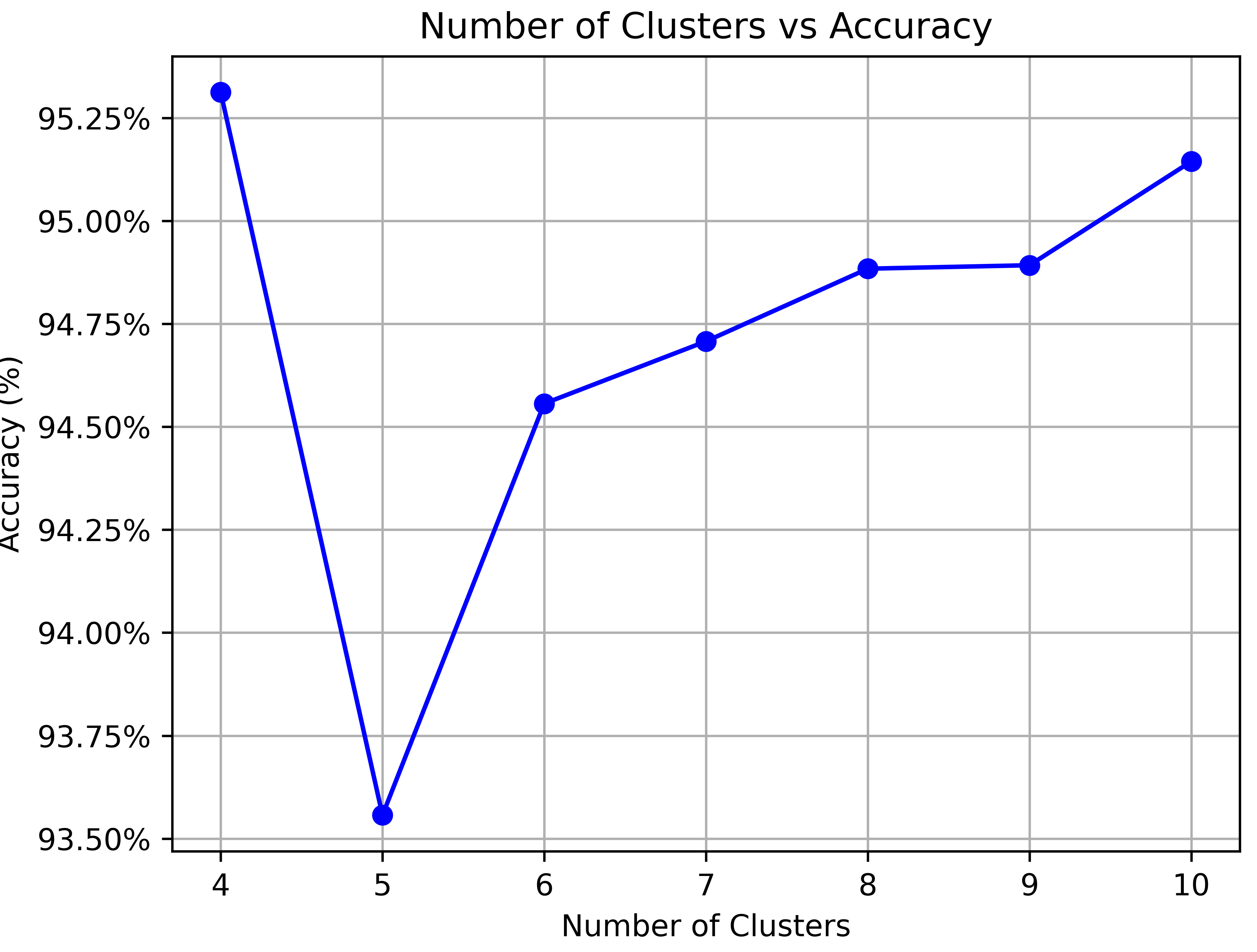}
    \caption{Clusters vs Accuracy}
    \label{fig:sub1_result_compare_clusters}
  \end{subfigure}
  \hfill
  \begin{subfigure}{0.49\textwidth}
    \includegraphics[height=0.6\textwidth, width=\linewidth]{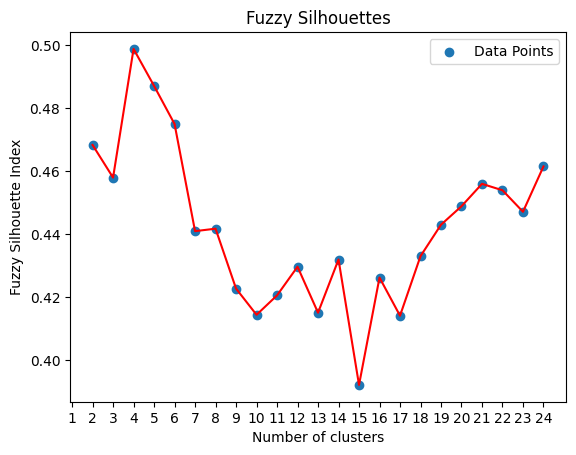}
    \caption{Silhouette Coefficient}
    \label{fig:sub2_result_compare_clusters}
  \end{subfigure}
\caption{Comparison of Accuracies With Model-2 Based on Fuzzy C-Means}
\label{fig:result_compare_clusters}
\end{figure*}

Model-1 is tested with type-2 fuzzy membership degrees and 24 emotion classes. Model-2, which is based on the unsupervised fuzzy clusters, we experimented with various numbers of clusters. The optimum number of clusters is 4 with the fuzzy Silhouette coefficient. We started with 4 clusters and ran the model up to 10 clusters. Results are shown in Fig.\ref{fig:result_compare_clusters}. Model-3 is based on the cuboid probabilistic lattice and also utilizes dual output. It receives the input of type-2 fuzzy membership degrees with 24 classes. All the results of the three models are shown in Table\ref{tab:accuracies}. Model-1 performed best with 96.09\% accuracy, followed by Model-3 with 95.75\% accuracy and the last, Model-2 with 95.31\% accuracy.

\begin{figure}
\centering
\includegraphics[width=\linewidth]{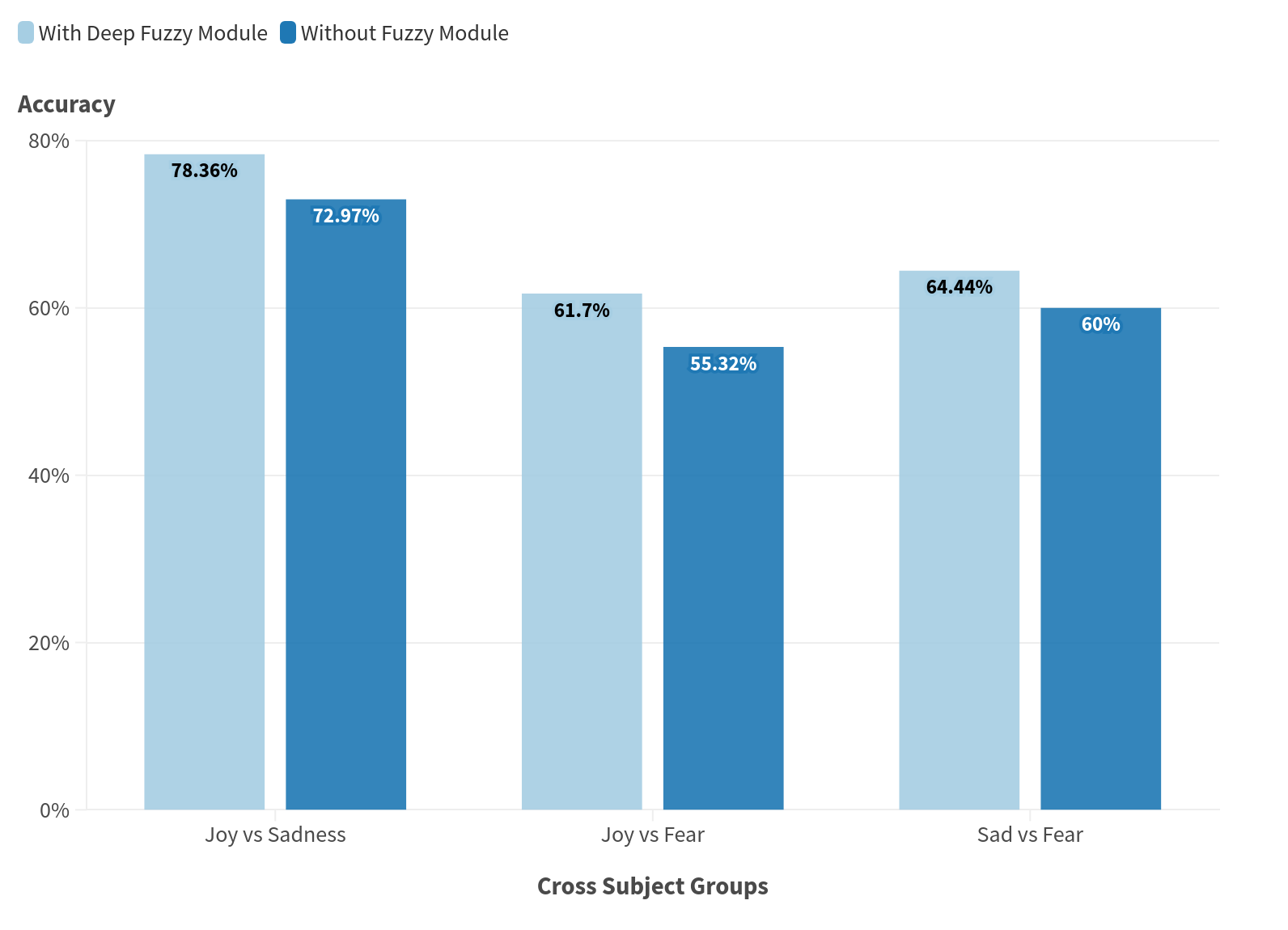}
\caption{Cross Subject Accuracy Comparison- with and without Deep Fuzzy Module}
\label{fig:cross_sub_comp}
\end{figure}

We also explored cross-subject emotion classification, which is considered a relatively more challenging task. Results are shown in Table\ref{tab:cross_sub_acc}. Results show that Group-1 vs Group-2 secures the best accuracy. Improvements when using a deep fuzzy framework are shown in Fig.\ref{fig:cross_sub_comp}.

An ablation study with many experiments is also conducted to check the performance of our proposed deep fuzzy framework. Firstly, we evaluated the performance of our models under various conditions without the fuzzy framework and only the module working as a crisp VAD space. secondly, we removed all the VAD space and recognised the emotions with only spatial and temporal modules. Thirdly, individual UMF and individual LMF working as type-1 fuzzy are evaluated to check the performance of type-2 fuzzy sets with a comparison to type-1 sets. All the results of ablation study are shown in Table\ref{tab:ablation}.

\begin{table}
\centering
\begin{tabular}{c|c|c|c}
\hline
\textbf{Models} & \textbf{Fuzzy Framework} & \textbf{\#Classes} & \textbf{Accuracy}\\
\hline
Model-1 & Type-2 Fuzzy Membership & 24 & 96.09\% \\
Model-2 & Unsupervised Fuzzy Clusters & 24 & 95.31\% \\
Model-3 & Cubical Probabilistic Lattice & 24 & 95.75\% \\
\hline
\end{tabular}
\caption{Results of All the Models with Various Fuzzy Representations.} \label{tab:accuracies}
\end{table}

\begin{table}
\centering
\begin{tabular}{c|c|c|c|c}
\hline
\textbf{Groups} & \textbf{Model} & \textbf{Framework} & \textbf{\#Classes} & \textbf{Accuracy}\\
\hline
G-1 vs G-2 & Model-1 & with fuzzy & 2 & 78.35\% \\
G-1 vs G-2 & Model-1 & without fuzzy & 2 & 72.97\% \\
\hline
G-1 vs G-3 & Model-1 & with fuzzy & 2 & 61.70\% \\
G-1 vs G-3 & Model-1 & without fuzzy & 2 & 55.31\% \\
\hline
G-2 vs G-3 & Model-1 & with fuzzy & 2 & 64.44\% \\
G-2 vs G-3 & Model-1 & without fuzzy & 2 & 60.00\% \\
\hline
\end{tabular}
\caption{Results of Cross-Subject Emotion Recognition.} 
\label{tab:cross_sub_acc}
\end{table}

\begin{table}
\centering
\begin{tabular}{m{1cm}|p{4cm}|c|c}
\hline
\textbf{Model} & \centering \textbf{Method} & \textbf{Classes} & \textbf{Accuracy}\\
\hline
Model-1 & without Fuzzy framework (with Crisp VAD values) & 24 & 95.01\%\\
\hline
Model-1 & without Fuzzy framework (without any VAD values) & 24 & 93.54\%\\
\hline
Model-1 & UMF working as Type-1 Fuzzy & 24 & 95.82\% \\
Model-1 & LMF working as Type-1 Fuzzy & 24 & 94.65\% \\
\hline
Model-3 & UMF working as Type-1 Fuzzy & 24 & 95.16\% \\
Model-3 & LMF working as Type-1 Fuzzy & 24 & 95.22\% \\
\hline
\end{tabular}
\caption{Results of Ablation Study}
\label{tab:ablation}
\end{table}

\section{Discussion and Conclusion}
The utilization of a fuzzy VAD space and the integration with the measured EEG data have helped in the development of a deep fuzzy framework for emotion representation and recognition. By introducing the fuzziness of type-2, the model becomes more adaptive to individual variations in emotional experience, enhancing its versatility across different experiments and diverse individual responses.

The DENS dataset, unlike others, has given an option to the subjects to choose the name and time of their emotional experience in the value of V,A and D. However, consideration of raw VAD values could be an oversimplification since mental activities may not be accurately represented using only three numbers. To address this issue, we fuzzified the VAD values and incorporated EEG (Electroencephalography) values in our deep fuzzy model. The process of fuzzifying VAD values improved the overall accuracy of the model in comparison to the base model.

The results shown are promising, and the highest accuracy achieved is 96.09\% with a type-2 fuzzy framework (Model-1) (Section-\ref{sec:model1})\ref{fig:models}. With the type-2 fuzzy framework, we have the leverage to take advantage of both the general representation of the emotion and the population-specific representation of the emotion if we use them as LMF and UMF. On the other hand, using type-1 representation, we obtained an accuracy of 95.82\% and 94.66\%, respectively, when we used these UMF and LMF separately working as type-1. This further supports our theory that this model works well as a generic model for emotion representation and recognition.

Cuboid probabilistic lattice-based representation (Model-3)(Section-\ref{sec:model3})\ref{fig:models} achieves an accuracy of 95.75\% when input is given by a type-2 membership degree. However, the accuracy decreases to 95.16\% and 95.22\%, respectively, when independently using UMF and LMF as type-1 membership degrees. Nevertheless, this cuboid-based probability lattice is able to enhance the performance almost to the level of when UMF is used as type-1 in the previous model. This is due to the fact that this model is dual-output based, and cuboid probabilistic lattices are optimized for the next input to the EEG features, thereby improving performance.

Finally, when it comes to unsupervised learning using a fuzzy C-means cluster-based model (Model-2)(Section-\ref{sec:model2})\ref{fig:models}, the model's accuracy tends to fluctuate depending on the number of clusters used. To determine the optimal number of clusters, we can use the fuzzy Silhouette coefficient. In this case, the coefficient suggests that four clusters is the optimal number. Beyond five clusters, the accuracy increases a bit but again drops.

We conducted an ablation study to evaluate the performance of our model in two scenarios: with no VAD space integrated and with crisp VAD space integrated into the architecture. The results showed that when no VAD space was used, the model's performance was as low as 93.54\%. However, with the crisp VAD space, the model's performance improved to 95.01\%. Hence, supporting the need for a type-2 fuzzy framework. Additionally, incorporating both physiological data (EEG features) and subjective ratings enhances the model's ability to capture the multidimensional aspects of emotions, leading to more reliable recognition outcomes.

In cross-subject analysis, our model achieves an accuracy of 78.37\%, which is also on par with the other recent works\cite{10209178, 10329776, xu2023subject}. Notably, our ablation study showed that changing no other parameters resulted in a significant improvement in accuracy from 72.97\% to 78.37\% (Fig.\ref{fig:cross_sub_comp}) if we add a fuzzy framework to the existing architecture. Generally, cross-subject studies do not yield significant accuracies\cite{10209178, lawhern2018eegnet, yin2020locally}. Also, a comparison of the other two groups shows a similar kind of improvement after adding the fuzzy framework to the existing architecture; however, they elicit a bit low accuracy but are still on par with the recent trends. This signifies the generalizability of the proposed architecture, showcasing its potential application across diverse circumstances. It is a critical step toward real-world applicability, and the study's success in this aspect further validates the robustness of the developed model. 

The study deals with 24 emotions at a time, which is not only challenging but also the first of its kind. Typically, emotion recognition tasks involve identifying emotional dimensions such as low or high valence and arousal. Additionally, they often focus on a few basic emotions, such as happiness, sadness, anger, and joy. The number of emotional classes is also kept low. In this perspective, the study's outcomes have significant implications for emotion recognition research and applications. The generic nature of the developed model, along with its successful cross-subject predictions, gives direction for real-world applications in the field of affective computing, human-computer interaction, mental health monitoring, etc. Future research could explore the adaptation of the model to different cultural contexts and the integration of additional modalities for an even more comprehensive understanding of human emotions, including complex emotions.

\section*{Acknowledgments}
Special thanks to Ms. Diya Srivastava, for her efforts in this work. Your assistance has been truly appreciated.

\bibliographystyle{IEEEtran}
\bibliography{bibliography}

\begin{IEEEbiography}[{\includegraphics[width=1in,height=1.25in, clip,keepaspectratio]{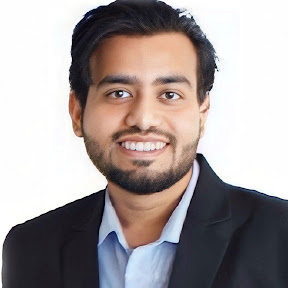}}]{Mohammad Asif} (Graduate Student Member, IEEE) received the master’s degree in Cognitive Science and also a master degree in Information Technology (specializing in software engineering) from Indian Institute of Information Technology, Allahabad. He holds a bachelor’s degree in computer science. 

He is currently a Research Scholar with the Indian Institute of Information Technology Allahabad, Prayagraj. His research interest includes Affective Computing. He is also working on Emotion Representation and Recognition using brain signals. He is using EEG for emotion detection using validated stimuli. He is also working on deep learning architectures for better AI advancement.

\end{IEEEbiography}

\begin{IEEEbiography}[{\includegraphics[width=1in,height=1.25in, clip,keepaspectratio]{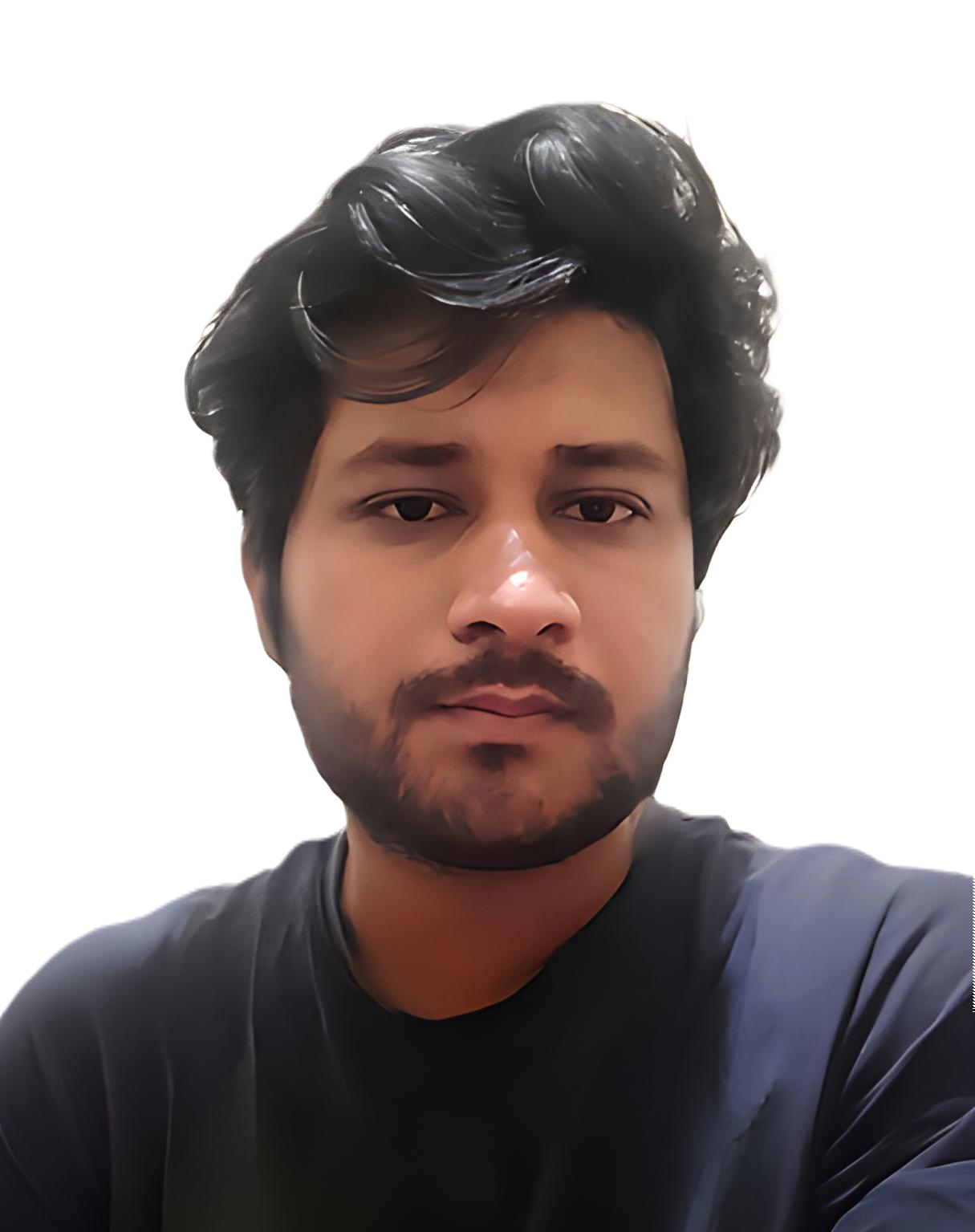}}]{Noman Ali}  is presently working toward a bachelors degree in Programming and Data Science with the Indian Institute of Technology Madras. Concurrently, he is also working towards a BE in Electronics and Communication Engineering from Panjab University. Noman's primary research interests encompass the dynamic fields of machine learning, deep learning, and computer vision. Additionally, he is actively engaged in developing machine learning applications in Earth Science.
\end{IEEEbiography}

\begin{IEEEbiography}[{\includegraphics[width=1in,height=1.25in, clip,keepaspectratio]{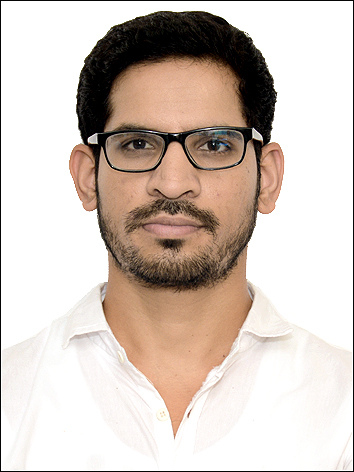}}]{Sudhakar Mishra} received his PhD from Indian Institute of Information Technology Allahabad, Prayagraj, India. His research work includes development of affective multimedia dataset validated on the Indian population. In addition, he has captured emotional events with brain recording using the Emotional Event Marker Paradigm. He is currently doing research on Buddhist analytical meditation and it's effect on emotional experience and emotion regulation skills. Currently, he is affiliated with IIT-Kanpur and University of Groningen. He is a member of the Society for Neuroscience.

\end{IEEEbiography}

\begin{IEEEbiography}[{\includegraphics[width=1in,height=1.25in, clip,keepaspectratio]{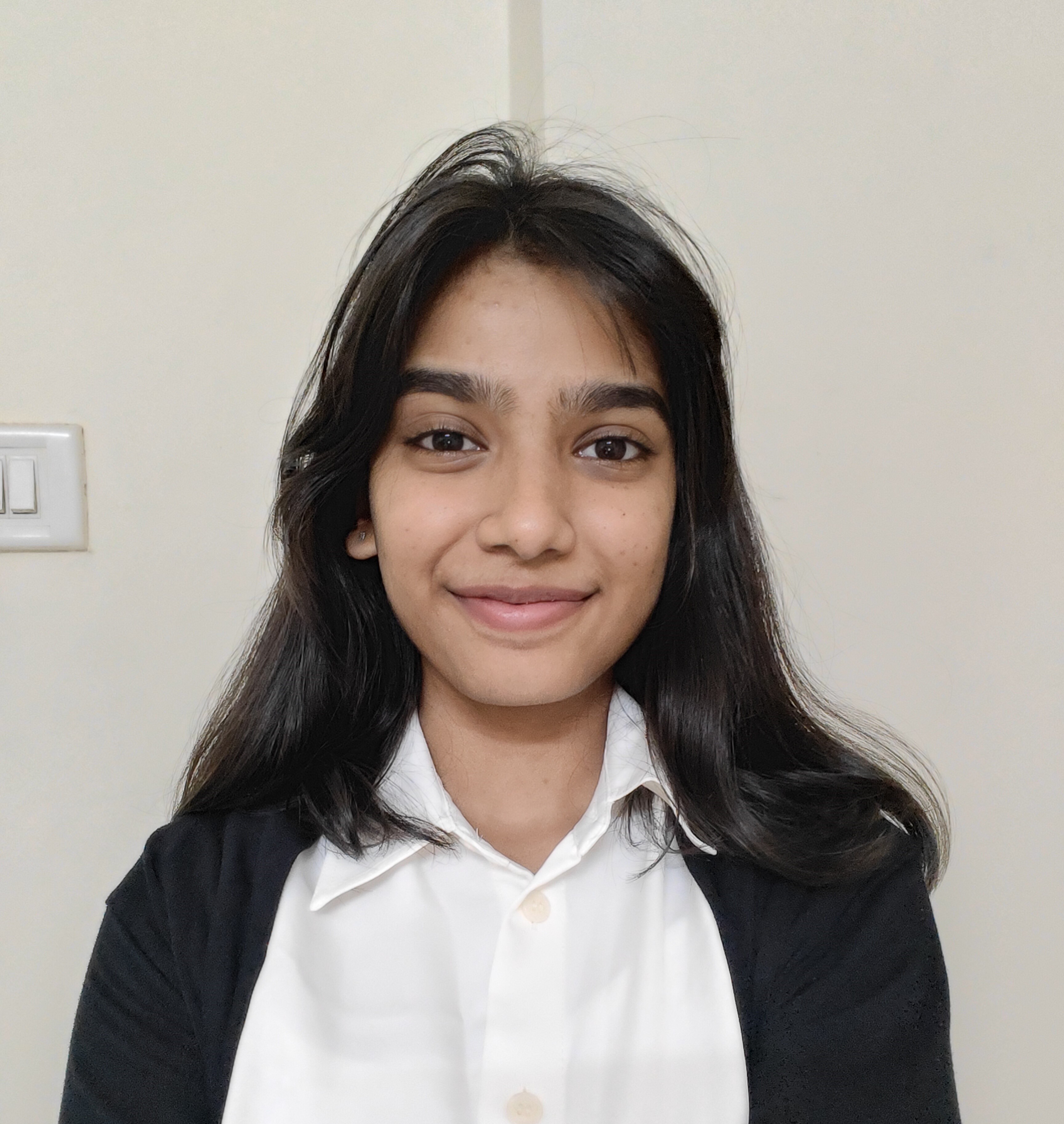}}]{Anushka Dandawate} is currently an undergraduate student pursuing a B.Tech in Information Technology at the Indian Institute of Information Technology, Allahabad. Actively exploring the intricacies of brain function using EEG signals, her research interests include deep learning, Cognitive science, and computational neuroscience. Her work contributes towards the intersection of technology and cognitive sciences.
\end{IEEEbiography}

\begin{IEEEbiography}[{\includegraphics[width=1in,height=1.25in, clip,keepaspectratio]{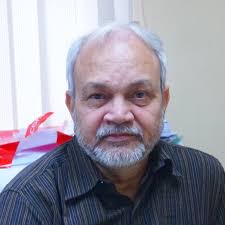}}]{Uma Shanker Tiwary}  (Senior Member, IEEE) received the Ph.D. degree from the Department of Electronics Engineering, Institute of Technology, Banaras Hindu University, Varanasi, India, in 1991. 

He is currently a professor in the Department of Information Technology, Dean (Faculty Affairs), and Head of the Centre for Cognitive Computing at the Indian Institute of Information Technology Allahabad. He holds research and teaching experience for more than 30 years, in which he is very much involved in image processing, computer vision, medical image processing, pattern recognition and script analysis, digital signal processing, speech and language processing, wavelet transforms, soft computing and fuzzy logic, neurocomputing and soft computers, speech-driven computers, natural language processing, brain simulation, cognitive science, and affective computing.
\end{IEEEbiography}



\vfill

\end{document}